\documentclass[12pt]{article}

\usepackage{multirow}
\usepackage{jheppub}

\usepackage{graphicx,psfrag, subfigure}
\usepackage{amsmath}

\usepackage{dsfont,caption}
\usepackage{setspace}
\usepackage{soul}

\numberwithin{equation}{section}
\usepackage{relsize}
\usepackage{hyperref}

\usepackage{amssymb}

\newcommand{\bs}{\boldsymbol}

\newcommand{\diag}{\text{diag}}

\newcommand{\di}{\mathrm{d}}

\DeclareMathAlphabet\mathbfcal{OMS}{cmsy}{b}{n}
\DeclareSymbolFont{bbold}{U}{bbold}{m}{n}

\DeclareSymbolFontAlphabet{\mathbbold}{bbold}
\DeclareSymbolFontAlphabet{\mathbbold}{bbold}

\makeatletter
\newcommand{\fixed@sra}{$\vrule height 2\fontdimen22\textfont2 width 0pt\shortrightarrow$}
\newcommand{\shortarrow}[1]{%
  \mathrel{\text{\rotatebox[origin=c]{\numexpr#1*45}{\fixed@sra}}}
}

\newcommand{\Po}{\bs P}
\newcommand{\Pob}{\bs  P^{\perp}}
\def\simleq{\; \raise0.3ex\hbox{$<$\kern-0.75em
      \raise-1.1ex\hbox{$\sim$}}\; }
\def\simgeq{\; \raise0.3ex\hbox{$>$\kern-0.75em
      \raise-1.1ex\hbox{$\sim$}}\; }

\def\M{M_{{\rm{Pl}}}}

\def\be{\begin{equation}}
\def\ee{\end{equation}}
\def\bea{\begin{eqnarray}}
\def\eea{\end{eqnarray}}

\def\be{\begin{equation}}
\def\ee{\end{equation}}
\def\bea{\begin{eqnarray}}
\def\eea{\end{eqnarray}}

\definecolor{mkcolor}{rgb}{1, .83,.83}
\definecolor{tbcolor}{rgb}{.83, .9,.83}
\definecolor{ojcolor}{rgb}{.83,.83, 1}
\definecolor{kecolor}{rgb}{1, 1, .5}

\begin{document}
\begin{titlepage}
\setcounter{page}{1} \baselineskip=15.5pt \thispagestyle{empty}
\bigskip\
\begin{center}

{\Large \bf Axion Landscape Cosmology}
\vskip 5pt
\vskip 15pt
\end{center}
\vspace{0.5cm}
\begin{center}
 
{Thomas C. Bachlechner$^*$, Kate Eckerle$^{ \dagger, \ddagger}$, Oliver Janssen$^\natural$ and Matthew Kleban$^\natural$}

\vspace{1cm}

\textsl{$^*$Department of Physics, University of California San Diego, La Jolla, USA} \\ \vspace{0.2cm}
\textsl{$^{ \dagger}$Dipartimento di Fisica, Universit\`a di Milano-Bicocca, Milan, Italy}\\ \vspace{0.2cm}
\textsl{$^\ddagger$INFN, sezione di Milano-Bicocca, Milan, Italy}\\ \vspace{0.2cm}
\textsl{$^\natural$Center for Cosmology and Particle Physics, New York University, New York, USA}
\end{center}

{\small  \noindent  \\[1cm]
\begin{center}
	\textbf{Abstract}
\end{center}
We study the cosmology of complex multi-axion  theories. With $\mathcal{O}(100)$ fields and GUT scale energies these theories contain a vast number of vacua, inflationary trajectories and a natural dark matter candidate. We demonstrate that the vacua are  stable on cosmological timescales. In a single theory,  both large- and small-field inflation are possible and yield a broad range of cosmological observables, and vacuum decay can be followed by a relatively large number ($> 60$) of efolds of inflation.  Light axions stabilized by gravitational instantons may constitute a natural dark matter candidate that does not spoil an axion solution to the strong CP problem.
\noindent}
\vspace{0.9cm}

\vfill
\begin{flushleft}
\small \today
\end{flushleft}
\end{titlepage}
\tableofcontents
\newpage

\section{Introduction}\label{intro}
The landscape paradigm for solving the cosmological constant (CC) problem requires the existence of an enormous number of meta-stable phases (``vacua") with differing vacuum energies \cite{Weinberg:1987dv,Bousso:2000xa}.  For theories where the fundamental scale is of order the Planck scale there must be $N_\text{vac} \simgeq M_\text{Pl}^4/\rho_\text{DE} \sim 10^{120}$ such local minima of the potential, where  $\rho_\text{DE}$ is the observed dark energy density.  Theories complex enough to contain such a large number of phases are in general extremely difficult to analyze at any level.  Unfortunately, the dynamics of the theory are essential to the putative solution of the CC problem.  The reason is that while small vacuum energy is necessary to allow structures to form, it is not sufficient.  The cosmological histories may be such that other effects prohibit structure formation.  The prototypical example of this is when the low-CC vacua are populated by tunneling, as one expects to be generic in the landscape.  In this case the negative curvature of the initial universe after the tunneling inhibits structure formation even when the CC is small \cite{Freivogel:2005vv}.  To avoid this, the tunneling must be followed by $\sim 60$ efolds of inflation (assuming a level of initial density perturbations roughly commensurate with observation).  Without inflation no structures form despite the small CC and the anthropic argument for the small CC fails.  But landscapes with small CC minima may not contain such  trajectories (for instance, the ``double well to the power $N$" toy landscape of \cite{ArkaniHamed:2005yv}).

In \cite{bejk1,bejk2} we developed a powerful framework for analyzing general theories involving $N$ axion fields $\theta^i$ coupled through a potential comprised of $P>N$ non-perturbative effects, and this is the third paper in this series. 
Our technique is based on identifying the set of  exact and approximate shift symmetries of the  axion potential. These symmetries are an extraordinarily powerful tool because the approximate symmetries are often extremely close to exact.  For instance, this renders the task of locating the potential's critical points tractable, even in field spaces with hundreds of dimensions.  Once equipped with the symmetries one may apply repeated shifts to the global minimum by the approximate symmetries, and mod out by the exact ones, to enumerate the distinct local minima.  
In addition to  the field space locations of extrema, we also showed how to retrieve important features of the potential like Hessian eigenvalues, and field ranges in the basins of attraction of minima from simple computations.  The purpose of this paper is to apply that formalism to the context of cosmology and to determine whether axion landscapes can solve the CC problem.

The techniques developed in \cite{bejk2} apply to axion potentials of the form
\begin{equation} \label{axion-potential}
V = V_0 + \sum_{I=1}^P \Lambda_I^4 \left[ 1-\cos(\mathbfcal{Q}\bs \theta + \bs \delta)^I \right].
\end{equation} 
Our notation is that of \cite{bejk2}: bold represents a vector or matrix.  Here  $V_0$ is a constant, the $\Lambda_I^4$ are the couplings of the axions to the  non-perturbative effects, $\bs \theta$ are the $N$ axion fields, $\mathbfcal{Q}$ is a $P\times N$ rank $N$ matrix containing integer charges $\mathcal{Q}^I_{\, j}$, and $\delta^I$ are constant phases.  When $ P > N$ and $P - N \ll N$ the $\delta^I$ can be set to zero to a very good approximation by a shift in field space \cite{bejk2}. The kinetic term is  assumed to be field-independent:
\begin{equation} \label{axionlagrangian}
\mathcal{L}_{\text{axion}}=\frac{1}{2}\partial \bs \theta^\top \bs K \partial\bs \theta-V \,,
\end{equation}
with $\bs K$  a positive definite $N \times N$ matrix.

The theory (\ref{axionlagrangian}) is motivated by the study of compactifications of string theory, where there are often hundreds axion fields \cite{Denef:2004cf,Douglas:2006es,Marsh:2011aa,Long:2014fba,Bachlechner:2014gfa,Long:2016jvd,Halverson:2017ffz,Halverson:2017deq,Demirtas:2018akl}.  The shift symmetry of the axions  is broken by non-perturbative effects, giving rise to a potential of the form (\ref{axion-potential}).  We model this by taking the charge matrix $\mathbfcal{Q}$ as a  random matrix  with independent identically distributed integer entries $\mathcal{Q}^I_{\, j}$, with variance $\sigma^2_\mathcal{Q}$. The simplest choice for the $\Lambda_{I}$ and $\bs K$  is $\Lambda_{I} = \Lambda$, $\forall I$, and $\bs K=f^2 \mathbbold{1}$ with fixed $f$. In \cite{bejk2} we considered much more general random (positive) ensembles.  {In most of this paper we will stick with the simplest choice, although in \S \ref{infdynamics} we discuss more general $\bs K$, and in \S \ref{lightaxionpheno} we consider the large hierarchy in the $\Lambda_I$ that can arise when some of the terms in (\ref{axion-potential}) come from gravitationally suppressed instantons.}

Because we are motivated by string theory, we will choose all the dimensionful parameters at the same scale: $\Lambda, f \sim M_\text{GUT} \sim M_\text{string} \sim 10^{-2} M_\text{Pl}$ and  $N \sim \text{few} \times 10^{2}$.  In the later sections of the paper we will consider a minimal coupling to QCD or a $U(1)$ with random, $M_\text{GUT}$-suppressed couplings, and the effects of gravitational instantons with actions of order $M_\text{Pl}/f$.  With these parameters and the techniques of \cite{bejk2} we can test whether this landscape truly solves the CC problem -- that is, whether those cosmological histories in which  collapsed structures form  (\` a la \cite{Weinberg:1987dv,Freivogel:2005vv}) do in fact resemble our universe.

Remarkably, without any model building and with only this simple requirement, typical cosmological histories have the following features our universe:
\begin{itemize}
\item{An extremely small CC, of order $\rho_\text{DE} \propto e^{-\mathcal{O}(1)\times {\M\over M_\text{GUT}}}$.}
\item{ An age of over $10$  billion years.}
\item{Approximately $60$ efolds of slow roll inflation with a primordial power spectrum $\delta \rho / \rho \sim 10^{-5}$.}
\item{Reheating following inflation. }
\item{Roughly the observed abundance of dark matter.}
\end{itemize}
In the final, upcoming work \cite{axidentaluniverse} of this series of papers we will demonstrate how the above features arise naturally in multi-axion theories simply from restricting to cosmological histories in which gravitationally collapsed structures can form.  For instance, the exponential suppression of the CC in the first bullet point originates from the small dark matter density relative to the  density of radiation, which itself arises naturally from ultra-light axions that  interact only with gravity.  In this paper we lay the groundwork for that analysis by studying the cosmology of multi-axion theories more generally, including observables such as the inflationary power spectrum, the abundance of dark matter and the status of an axion solution to the strong CP problem.

The structure of this paper follows the order of our list. In \S \ref{sec:vactransitions} we consider vacuum decay and demonstrate that a vast number of vacua are stable on cosmological timescales. In \S \ref{infdynamics} we discuss how the inflationary dynamics and observables in extremely complex multi-axion theories can be sampled efficiently, and we demonstrate that a single theory allows for a broad range of observables. We discuss fuzzy dark matter in \S\ref{lightaxionpheno} and show that gravitational instantons do not spoil the axion solution of the strong CP problem.

\section{Vacuum transitions} \label{sec:vactransitions}
In the semiclassical approximation to the decay of de Sitter vacua, there are two mechanisms at work\footnote{TB does not concur with the results of \cite{Batra:2006rz,Coleman:1980aw,Hawking:1982my} regarding vacuum transition rates in gravitational theories \cite{Bachlechner:2018pqk,Bachlechner:2018jmq}.}:  quantum tunneling through a barrier and thermal evaporation to the top of a barrier \cite{Batra:2006rz}. The decay proceeds either by a Coleman-de Luccia (CdL) instanton \cite{Coleman:1980aw} that represents the least-action combination of these two mechanisms, or solely by thermal evaporation via a Hawking-Moss (HM) instanton \cite{Hawking:1982my}.  

In this section we examine decays for a certain class of vacua in our axion landscape, namely those with vacuum energy density sufficiently close to the global minimum that a quadratic approximation of the potential is applicable. Specifically, this means we will focus on vacua for which the arguments of the cosines, $(\mathbfcal{Q}\bs \theta)^I $, in (\ref{axion-potential}) are close to integer multiples of $2\pi$ (for all $1 \leq I \leq P$). Such vacua are under very good analytic control. We will avoid a precise definition of the quadratic domain, i.e. which vacua we consider to be well-described by a quadratic expansion of the potential, as our  qualitative results are independent of the precise choice. 

When $-V_0 = |V_0| \ll \Lambda^4$, all vacua with nearly zero or negative vacuum energy are guaranteed to fall into the quadratic domain, because if the argument of any cosine is substantially different from zero (mod $2 \pi$) its positive contribution to the potential (\ref{axion-potential}) renders the total potential energy of order $+\Lambda^4 \gg 0$. The size of the hierarchy required between $|V_0|$ and $\Lambda^4$ depends on the desired accuracy of the quadratic expansion.  A factor of a few suffices to estimate the decay exponents (the instanton action) to ${\cal{O}}(1)$ accuracy. Therefore, at least if $ 0 < -V_0 \ll \Lambda^4$ this quadratic approximation suffices for purposes of studying vacua with small vacuum energy, and for studying the decay of such vacua to any lower energy minima. 

We will see that in the parameter regime we are focusing on, CdL decays are the dominant channel and vacua with small CC are typically long-lived on cosmological timescales. However the quadratic approximation does not suffice for studying all decays \emph{into} such vacua, as those can originate from higher regions of the potential that might not fall into the quadratic domain.  For vacua in the quadratic domain, we find an upper bound on the decay rate, and conclude that no significant fine-tuning is necessary for the vacua to be meta-stable on cosmological timescales.

\subsection{Hawking-Moss decays} \label{HM-section}
HM instantons are configurations in Euclidean signature de Sitter space where the field is constant at a saddle point of the potential. These instantons are potentially relevant for decay only if the saddle has degree $k=1$, meaning the Hessian evaluated there has exactly one negative eigenvalue  $V''_* < 0$, and if  
\begin{equation} \label{HMcondition}
	M_{\text{Pl}}^2 \frac{|V''_*|}{V_*} \leq \frac{4}{3} \,,
\end{equation}
where $V_*$ is the potential energy at the saddle point supporting the instanton interpolating between the ``parent'' and ``target'' vacua (see e.g.~\cite{Batra:2006rz}). If  these conditions are satisfied the instanton has a single negative mode and contributes an imaginary part to the energy and hence to the decay of the state. For saddle points where (\ref{HMcondition}) is not satisfied or there is more than one negative direction ($k > 1$), the HM instanton always has multiple negative modes and presumably does not contribute to the decay. In such cases a CdL instanton always exists \cite{Batra:2006rz}.

It is simple to estimate whether (\ref{HMcondition}) typically holds using the results of \cite{bejk2} (in particular \S 3.6.2) and appendix \ref{appendix-saddles} in this paper.  The analysis there shows that the $k=1$ saddles adjacent to minima in the quadratic domain have $V_* \approx 2\Lambda^4$  because one cosine reaches its maximum roughly halfway in between while the rest are constant.  The negative direction at the saddle satisfies $\langle |V_*''| \rangle \sim N \sigma^2_{\mathcal{Q}}\Lambda^4/ 2 f^2 \propto N$, with a standard deviation that  scales only as $\sqrt{N}$.  We are most interested in the rough parameter regime where $\sigma_\mathcal{Q} = \mathcal{O}(1), N \approx P \gg 1$ and $f \ll M_{\text{Pl}}$, such that with (\ref{HMcondition}) HM decays are typically suppressed,
\begin{equation}
	\langle M_{\text{Pl}}^2 \frac{|V''_*|}{V_*} \rangle \propto N \sigma_\mathcal{Q}^2 \left( \frac{M_{\text{Pl}}}{f} \right)^2 \gg 1 \,.
\end{equation}
 Note that even for $\sigma_\mathcal{Q} = \mathcal{O}(1/\sqrt{N})$, which one may also be interested in (e.g. \cite{Bachlechner:2014gfa,bejk2}), this inequality is satisfied. We conclude HM instantons are irrelevant for the decay of quadratic domain minima.

\subsection{Coleman-de Luccia decays}\label{stability-subsec}
When the  condition (\ref{HMcondition}) for HM decays is not satisfied,  decays will proceed via CdL transitions. We are mainly interested in studying the decay of vacua with small vacuum energy, those that can contain collapsed structures like galaxies. By Weinberg's famous anthropic argument \cite{Weinberg:1987dv} these have $|V_{\text{vac}}| \lesssim 10^{-120} M_\text{Pl}^4$. This very narrow band is nevertheless densely populated in the landscapes we are studying \cite{bejk1,bejk2}. For $|V_0| \ll \Lambda^4$, all such vacua are in the quadratic domain.  As mentioned above this ensures the validity of the approximations we make to certain characteristics of the potential such as the location of critical points, their heights and their Hessian eigenvalues.

In general little is rigorously known about tunneling in high-dimensional landscapes, especially when the effects of gravity are included. For flat space tunneling  more is known. For instance, we can assume the dominant instanton has maximal spherical symmetry \cite{Coleman:1977th,Blum:2016ipp}. In the following, we will use the thin-wall approximation to bound the decay rates. Thin-wall should be justified when (\ref{HMcondition}) is strongly violated, and  the numerical checks described in \S \ref{gradflow} support this conclusion.  

\subsubsection{Neighboring minima} \label{neighborsec}
A given (quadratic domain) vacuum can tunnel into any neighboring vacuum that has lower energy.\footnote{There may be decay channels to minima even further away, but these are presumably suppressed. We only consider tunneling to lower energy vacua in this paper. Upward transitions from de Sitter minima are not impossible, but are exponentially suppressed.} A method that accurately locates the neighboring minima can be deduced from some considerations regarding the potential (\ref{axion-potential}). $V$ is invariant under shifts of the  arguments of the cosines by $2 \pi \bs{v}_k$, where $\bs{v}_k$ is an integer $P$-vector with $k$ non-vanishing  components of $\pm 1$. We call a degree-$k$ neighbor a vacuum that is displaced from another vacuum by a shift in the $N$-dimensional axion field space, such that $k$ cosine arguments shift by roughly $2\pi$, i.e.
\be
\mathbfcal Q\bs \theta_\text{neighbor}=\mathbfcal Q\bs \theta_\text{vacuum}+2 \pi \bs{v}_k \,.
\ee
Each vacuum has at most $3^P-1$ neighboring vacua.  However, since the argument of each cosine is a linear combination of the $N$ axion fields,  shifting the cosine arguments this way requires solving a set of $P$ linear equations in $N$ variables. When $P \leq N$ a solution always exists. When $P > N$ there are more equations than variables and such shifts (in general) do not exist. Nonetheless, when $P, N \gg P-N$ one can solve these equations approximately.  This shows that low-lying vacua have roughly $3^P$ neighbors \cite{bejk2}.  A priori, the decay could proceed by tunneling to any of this huge number of neighboring vacua. A special class are those separated from the decaying vacuum by shifting the argument of a \emph{single} cosine by $\pm 2 \pi$. There are  $2P$ such $k=1$ neighbors, some fraction of which have lower vacuum energy density than the decaying vacuum.\footnote{The amount depends on the height of the decaying minimum above the global minimum. The number will be small for very low-lying minima, while for higher-lying minima (but still in the quadratic domain) it is well-approximated by $P$.} We refer to these $k=1$ neighbors as ``face neighbors" (because they are separated from the decaying vacuum by a  face of a cube in the auxiliary field space defined in \cite{bejk2}).

Each face neighbor minimum is separated from the decaying vacuum by a barrier of height approximately $ 2 \Lambda^4$. The top of the barrier is generically a degree $k=1$ saddle point. Neighbors where $k$ cosines shift by $2 \pi$ are typically separated from the decaying vacuum by a barrier with height approximately $2 k \Lambda^4$, and by a degree-$k$ saddle point. We will call these ``degree-$k$ neighbors".  The typical distance to a degree-$k$ neighbor scales as $\sqrt{k}$ (due to a famous result of Pythagoras). Hence in addition to being the set of minima separated from the parent vacuum by the lowest barriers, the face neighbors are also those typically located within the shortest distance. This makes it plausible that the dominant decay channel will be to a face neighbor.

\subsubsection{Thin-wall tension} \label{TWtensionsec}
We write the semiclassical bubble nucleation rate per unit four-volume as
\begin{equation} \label{rateeq}
	\Gamma \sim A \, e^{-B} \,.
\end{equation}
For a single scalar in flat space, and in the thin-wall approximation, we have \cite{Coleman:1977py}
\begin{equation}
	B_\text{flat} \sim \frac{27 \pi^2}{2}\frac{\sigma^4}{\epsilon^3} \,,
\end{equation}
where $\epsilon = V_\text{max} - V_\text{min}$ is the difference in energy density between the two vacua and $\sigma$ is the tension of the bubble wall,
\begin{equation} \label{eq-tension}
\sigma = \int_{\varphi_\text{min}}^{\varphi_\text{max}} \hspace{-0.2cm} \di \varphi \, \sqrt{2(V(\varphi)-V_{\text{min}})} \,.
\end{equation}
($\varphi_\text{min,max}$ denote the locations of the lower-lying and higher-lying vacuum respectively.) By introducing a minimal wall tension, 
\begin{equation} \sigma_\text{min} \equiv \int_{\varphi_0}^{\varphi_\text{max}} \hspace{-0.1cm} \di \varphi \sqrt{2(V(\varphi)-V_{\text{max}})} \,, 
\end{equation} 
where $\varphi_0$ is defined by $V(\varphi_0) = V_\text{max}$, the thin-wall formula turns into a lower bound for $B_\text{flat}$ \cite{Brown:2017cca}:
\begin{equation} \label{Brownbound}
	B_\text{flat} \geq \frac{27 \pi^2}{2} \frac{\sigma_\text{min}^4}{\epsilon^3} \,.
\end{equation}
This inequality holds for any (single) scalar field theory -- even those for which the thin-wall approximation is not valid. Even if gravitational effects are not negligible, they only serve to increase $B$ for thin-wall tunneling from flat (or nearly flat) spacetime to AdS, which is the case at hand.  Finally, we have studied the instantons numerically and found that the thin-wall approximation does seem accurate in the regime we are focusing on (see also \S \ref{inflationaftertunneling}).

One can now use (\ref{eq-tension}) to estimate a lower bound on $B$, for a decay to a minimum separated from the parent by shifts of $k$ cosines by approximately $2 \pi$.  For the reason discussed above the height of the saddle point along such a direction is $V_* \sim k \, 2 \Lambda^4$, with second derivative $|V''_*| \approx N \sigma_\mathcal{Q}^2 \Lambda^4 / 2 f^2$. The typical field space distance across the barrier is $\Delta \varphi \sim 2 \sqrt{2 V_*/|V''_*|} \times \pi/4 \sim \pi \left( f / \sigma_\mathcal{Q} \right) \sqrt{2 k/N}$ (where we've approximated the barrier as a parabola), which gives
\begin{equation}
	\sigma_\text{min} \gtrsim \Delta \varphi \times \sqrt{2 V_*} \sim 2 \pi k \sqrt{\frac{2}{N}} \frac{f \Lambda^2}{\sigma_\mathcal{Q}} \,.
\end{equation}
The energy difference $\epsilon$ between a zero energy vacuum and one with negative energy  cannot exceed $V_0$ (the energy of the global minimum);
$\epsilon < |V_0| \ll \Lambda^4$. 

Using these estimates in  (\ref{Brownbound}) gives
\begin{equation} \label{boundsummary}
	B \gtrsim \frac{27 \pi^2}{2} \frac{\sigma_\text{min}^4}{\epsilon^3} \gtrsim \left({900 \over N}\right)^2 \times \left( {\Lambda^4 \over V_0 }\right)^3 \times \left( \frac{ f}{\sigma_\mathcal{Q} \Lambda} \right)^4 \times k^4 \,,
\end{equation}
which suggests that decays to face neighbors with $k=1$ are dominant. Thus, even with $N \approx 10^3$, with  $\Lambda < f$ and/or $V_0 < \Lambda^4$ one can achieve $B \gg 1$. Stability on the order of $10^{10}$ years requires $ B \gtrsim 10^3$ (since $4 \log {10^{10} \text{years} \over t_\text{Pl}} \approx 10^3$).  We have suppressed several steps in this analysis to give the reader the option of bypassing technical details if they wish. A thorough derivation can be found in appendices \ref{appendix-saddles} and \ref{appendix-stability}. We also include an analysis of the distribution of vacuum energy differences across the sets of degree-$k$ neighbors in section \ref{appendix-deltaVs}.  
  
\subsubsection{Numerical checks and the ``gradient flow approximation''} \label{gradflow}
It is difficult to check these approximations numerically due to the high dimension of the field space.  Even in field theories without gravity, to our knowledge the best current codes for studying vacuum decay can only handle roughly $N=5$ field space dimensions \cite{Masoumi:2016wot}.  However, the tools developed in \cite{bejk2} make a semi-analytic check available for axion theories.  

As we described above, out of $\sim 3^{P}$ neighboring vacua the  $2 P$  ``face neighbor'' channels are likely to dominate  the decay rate.  For those decays (or any degree-$k$ neighbor, in general) we can use our techniques to locate the lowest saddle point that separates the parent from the target.  We  then find the gradient line that connects the parent to the target and passes through this saddle point.  The  potential along this line can then be treated as if it were the potential for a single scalar, allowing us to numerically compute the CdL instanton.   We refer to this as the ``gradient flow approximation.''  It is approximate because the exact instanton does not necessarily follow the gradient flow, but we expect this method to correctly compute the instanton action and trajectory up to ${\cal{O}}(1)$ corrections.   The instanton computes the decay rate and provides the initial conditions for the cosmological dynamics after the tunneling, which we calculate using the full $N$-dimensional potential (see \S \ref{inflationaftertunneling}).

We  verified that the gradient flow indeed approximates the numerical results of \cite{Masoumi:2016wot} in the ``sum of cosines'' example considered there, and agrees with the results of the thin-wall analytic approximation we turn to next.   At least for our class of potentials, it may be an improvement over the ``straight line'' approximation introduced in \cite{Masoumi:2017trx}.

\section{Inflation}\label{infdynamics}
We now turn the topic of  inflationary dynamics. Much of the discussion in this section applies to general multi-axion theories, but we will focus particular attention on well-aligned theories (which are generic when $N \approx P \gg 1$) where it is a very good approximation to set the phases in the non-perturbative axion potential to zero  (see \cite{bejk1} for details).

Let us begin by recalling the action relevant for the inflationary dynamics driven by $N$ canonically normalized axions $	{\bs \Theta \equiv \sqrt{\bs K} \bs \theta}$,
\be
S=\int d^4 x \sqrt{-g} \left({\M^2\over 2}{R} -{1\over 2}\partial\bs\Theta^\top \partial\bs\Theta-V_{\text{axion}}(\bs\Theta)-V_0 \right)\,.
\ee
Here $g_{\mu \nu}$ is the flat FLRW metric with scale factor $a(t)$,
\be
\di s^2 = - \di t^2 + a(t)^2 \di \bs x^2\,.
\ee
As above we choose $V_0$ so that the axion contribution to the potential is  non-negative, with its global minimum at zero:
\be
V_{\text{axion}}=\sum_{I=1}^P \Lambda_I^4 \left[ 1 - \cos\left( \bold Q \bs \Theta \right)^I \right]\,,
\ee
where $\bold Q \equiv \mathbfcal Q \, \bs K^{-1/2}$ is the charge matrix for the canonically normalized fields $\bs \Theta$.

The equations of motion for the scale factor and axions are
\bea\label{eominflation}
(\Theta^i)'' &=&(\epsilon-3)(\Theta^i)' - {1\over H^2}{\partial V_{\text{axion}} \over \partial \Theta^i} ( \bs \Theta ) \,, \nonumber\\3\M^2 H^2&=&{V_{\text{axion}}( \bs \Theta)+V_0\over 1-\epsilon/3}\,,
\eea
where $' \equiv \di / \di N_e$, $N_e= \log \, a$ denotes the number of efolds, $H = (\di a / \di t) / a \equiv \dot{a} / a$ is the Hubble scale and the Hubble slow roll parameter $\epsilon$ is defined by
\be
\epsilon=- {\dot{H} \over H^{2}} = -{H' \over H} \,.
\ee

In general the dynamics of this system are quite complicated. However, the evolution may effectively be that of a single field if no isocurvature perturbations are sourced.  To make this manifest, following \cite{GrootNibbelink:2001qt,McAllister:2012am}, we decompose the fields into a basis  defined by unit vectors $\{\bs E_i\}$ along the inflationary trajectory:
\be
\bs E_i = {\bs P^\perp_{i-1} \bs \Theta^{(i)} \over \lVert \bs P^\perp_{i-1} \bs \Theta^{(i)} \rVert_2}\,,~~~~~\bs P^\perp_i= \mathbbold{1} - \sum_{j=1}^i \bs E_j \otimes\bs E_j \,,
\ee
where $i=1,\dots,N$, $\bs P^\perp_0 = \mathbbold{1}$ and $\bs P^\perp_i$ is a projection operator onto the subspace perpendicular to $\langle \bs E_1, \bs E_2, \dots, \bs E_i \rangle$. This decomposition of the field is very convenient: $\bs E_1$ corresponds to the instantaneous direction of the field velocity, while $\bs E_2$ indicates the direction of the acceleration transverse to the field velocity and signals multifield behavior. With this basis in mind we can decompose the second slow roll parameter $\bs\eta$ into components parallel and perpendicular to the field trajectory,
\be
\bs\eta={1\over H}{\ddot{\bs\Theta} \over \lVert \dot{\bs\Theta} \rVert_2} = {{\bs \Theta'' } - \epsilon \, {\bs \Theta'}\over \lVert \bs\Theta' \rVert_2}\,,~~~~~\eta_\parallel=\bs\eta \cdot\bs E_1\,,~~~~~\eta_\perp=\bs\eta \cdot\bs E_2\,.
\ee
This decomposition is particularly well-suited to study perturbations. Curvature (adiabatic) perturbations are described by perturbations in the direction of $\bs E_1$ and are the only relevant perturbations for single field inflation. In the case of single field inflation the curvature perturbations can immediately be related to density perturbations. Isocurvature (entropy) perturbations describe the relative decomposition of the energy density into the different field components. In the basis we have chosen, one isocurvature mode is distinguished in that it is the only one that couples to the adiabatic perturbation. This coupling is proportional to $\eta_\perp$, such that single field behavior is recovered when the trajectory does not turn in field space, i.e. $\eta_\perp=0$. 

Given an initial condition, we will  be concerned with solving the classical equations of motion and evaluating some basic observables, such as the spectral index and the tensor-to-scalar ratio,  using the leading expressions in the slow roll regime. The transverse slow roll parameter $\eta_\perp$ provides us with some  information about the consistency of the single field, slow roll approximation. 

We can define effective slow roll parameters by differentiating the potential along the inflationary trajectory.  These ``potential slow roll parameters'' are related to those defined above by
\be
\epsilon_V \equiv {\M^2\over 2} \left({\partial_{\Theta} V\over V}\right)^2\approx \epsilon\,,~~~~~\eta_V \equiv { \M^2}{\partial^2_{\Theta} V\over V}\approx \epsilon-\eta_\parallel\,,
\ee
where $\partial_{\Theta}$ denotes differentiation in the direction of the field space velocity $\bs E_{1}$, and the  approximation is valid in slow roll. In the single field, slow roll approximation where $\epsilon$ and $\sqrt{\epsilon} \, \eta_\parallel$ are small and $\eta_\perp$ vanishes throughout the inflationary evolution, the spectral index and tensor-to-scalar ratio are given by
\be
n_\text{s}\approx 1-2\eta_\parallel-4\epsilon\,,~~~~~r\approx 16\epsilon\,,
\ee
which are evaluated at the time when the CMB modes exit the horizon.  The amplitude of scalar temperature anisotropies is 
\be \label{droverr}
A_\text{s} \equiv {1 \over 24 \pi^2} {V \over \epsilon \M^4} \,,
\ee
with observed values of $A_\text{s} \approx 2.1 \times 10^{-9}$, $n_\text{s} \approx .965$, and $r \simleq .07$ \cite{Aghanim:2018eyx}. Using this value for $A_\text{s}$, the scale of inflation is related to the tensor-to-scalar ratio by (see e.g. \cite{Baumann:2014nda})
\be
V_{\text{inf}}\approx 3.2\times10^{-8}~r\M^4 \,.
\ee

 Near low-lying minima the axion potential is approximately quadratic.  For inflation in the quadratic regime near such minima, the first requirement is that the field range be long enough to produce a sufficient number of efolds of inflation.  In \cite{bejk2}, we estimated the ``diameter" of the region surrounding a typical minimum to be 
 \be\label{d0estimate}
{\cal D} \approx 2\pi \sqrt{P} {f\over \sigma_{\mathcal Q}}{1\over \sqrt{P} (1 - \sqrt{N/P})} \approx  2\pi \sqrt{P} {f\over \sigma_{\mathcal Q}} \frac{2 \sqrt{N}}{P - N} \,,
\ee
where the last approximate equality is valid for $P - N \ll N$.  To attain 60 efolds of inflation requires ${\cal D} \simgeq 10 \M$, which is possible with e.g.~$f\sim 10^{-2} \M, N \approx P \approx 500, \sigma_\mathcal{Q} \approx 1$.

For inflation taking place in this approximately quadratic regime around a low-lying minimum, the scale of the potential will be of order $\Lambda^4$ and the slow roll parameter $\epsilon$ will take values of order those for a quadratic ($\epsilon = 1/2 N_e$) or linear ($\epsilon = 1/4 N_e$) potential.  Assuming a roughly linear potential, the amplitude of the observed perturbations  will be approximately
\be \label{deltarhoquad}
A_\text{s} \approx {1 \over 24 \pi^2} \left( {\Lambda \over \M}\right)^4 \times 2 N_e \approx \left( {\Lambda \over \M}\right)^4 \, .
\ee

For GUT scale $\Lambda \sim 10^{-2} \M$, this gives roughly the observed amplitude, with tensors at the observed upper bound.  However, as we will see multi-axion theories contain many other types of inflationary trajectories, some of which are far from quadratic, or indeed from those of any standard inflationary potential. 

In \S\ref{samp} we discuss inflation  generally in axion landscapes.  In \S \ref{infexample} we numerically analyze inflation in a specific example, where we sample the landscape by choosing the starting point uniformly randomly in the field space (with zero initial velocity).  In \S\ref{inflationaftertunneling} we  discuss inflation where the initial conditions are set by tunneling from a higher minimum.   In \S \ref{reheat} we briefly sketch reheating when the axions are coupled to a gauge field.

\subsection{Sampling the theory}\label{samp}
{One could  study the inflationary dynamics in multi-axion theories, or random ensembles of such theories, in some generality by sampling over the parameters of the Lagrangian (\ref{axionlagrangian}).  These consist of the} metric on moduli space $\bs K$, the charge matrix $\mathbfcal Q$ and the global minimum of the vacuum energy density $V_0$.  The gravitational contributions to the axion potential will be irrelevant for the dynamics, but they may vastly increase the possible discrete vacuum energy densities in the theory, so we will assume that $V_0$ can be tuned to arbitrary accuracy. A choice for the parameter ensembles that is loosely inspired by explicit compactifications of string theory \cite{Denef:2004cf,Douglas:2006es,Marsh:2011aa,Long:2014fba,Bachlechner:2014gfa,Long:2016jvd,Halverson:2017ffz,Halverson:2017deq,Demirtas:2018akl} is as follows 
\begin{enumerate}
\item The metric $\bs K$ is a positive definite random matrix (for instance a Wishart or inverse Wishart matrix) with largest eigenvalue $f_N^2\lesssim \M^2$. 
\item The axion charge matrix $\mathbfcal{Q}$ is a sparse matrix of i.i.d.~random integers with a fraction $\gtrsim 3/N$ of non-vanishing entries.
\item The background vacuum energy density $V_0$, {uniformly distributed between $\pm \M^4$.}
\end{enumerate}

Even after fixing an ensemble of effective theories or even a unique theory, significant uncertainty remains due to the unknown weight with which different cosmological histories contribute to the distributions of observables -- i.e., the measure problem of inflationary cosmology. For example, there may exist a significant selection bias towards small final vacuum energy densities and sufficient inflation. To at least partially account for these selection biases {we only retain inflationary trajectories that satisfy the following:}

\begin{enumerate}
\item The vacuum energy density in the  minimum the trajectory ends in is not substantially larger than the observed dark energy density in our universe \cite{Weinberg:1987dv}.
\item Inflation lasts long enough to solve the horizon and flatness problems, which here for simplicity we take to mean that $N_e \geq 60$.
\end{enumerate}
In the related work \cite{axidentaluniverse} we demonstrate how these assumptions follow from the single requirement of structure formation.

\begin{figure}
  \centering
  \includegraphics[width=1\textwidth]{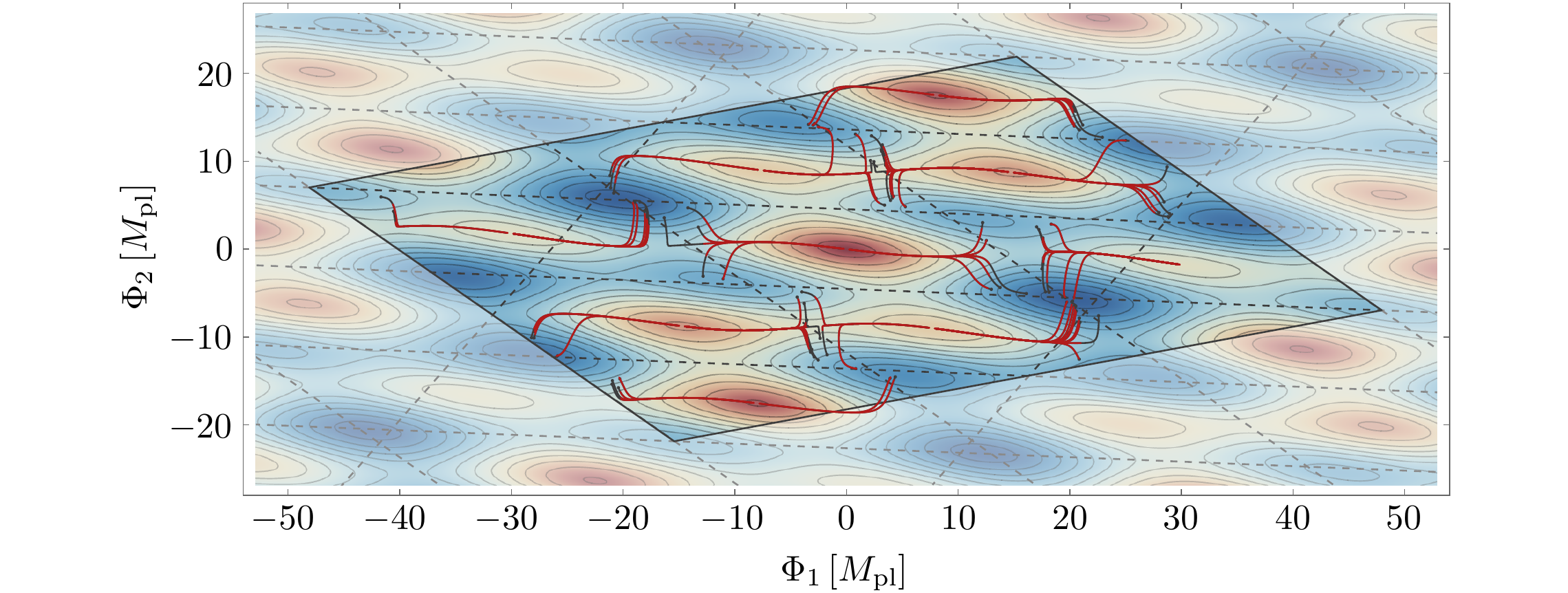}
  \caption{\small Complete sample of inflationary trajectories with more than $60$ efolds terminating at low-CC vacua for theory defined in (\ref{samplethry}). The full trajectories are shown in gray, while the last 60 efolds are colored red. The contour plot represents the axion potential. The periodic domain is highlighted and surrounded by gray solid lines. The boundaries of all tiles are denoted by dashed lines.}\label{nis3example}
\end{figure}

Since we assume that the background vacuum energy density is roughly uniformly distributed, any of the vacua of the axion theory may correspond to the cosmologically relevant late time vacuum.  We will reject any dynamics that do not terminate in a vacuum with vanishingly small vacuum energy density, so we can obtain a representative sample of inflationary observables by picking a representative sample of vacua at energy densities $V_{\text{vac},i}$ and then setting $V_0=-V_{\text{vac},i}$. For each $i$ we then choose initial conditions that are uniformly distributed over the periodic domain of the axion potential, and reject any trajectory that gives rise to less than $60$ efolds of inflation, or does not terminate at a vacuum with vanishing energy density, consistent with our assumptions about selection bias. In fact, it is not necessary to sample all vacua, nor to consider initial conditions uniformly distributed over the entire periodic domain of the potential. Merely considering the attractor regions in the vicinity of a representative sample of potential late time vacua provides a representative sample of the inflationary dynamics. This allows for a systematic study of potentials with exponentially many distinct vacua. We illustrate some possible inflationary trajectories for the specific potential discussed in the next section in Figure \ref{nis3example}. 

Now that we discussed how to systematically sample ensembles of axion theories we might embark on a detailed study of the distribution of inflationary observables. However, as long as the measure-dependent weight of each inflationary trajectory is unknown such a study is tentative to some extent, as we  cannot make {definite} predictions for cosmological observables.  Still, it may  be instructive to sample the inflationary dynamics. A comprehensive study of this kind is beyond the scope of this work, so in the following section we simply consider one particular axion theory.

\subsection{An explicit example}\label{infexample}
\begin{figure}
  \centering
  \includegraphics[width=1\textwidth]{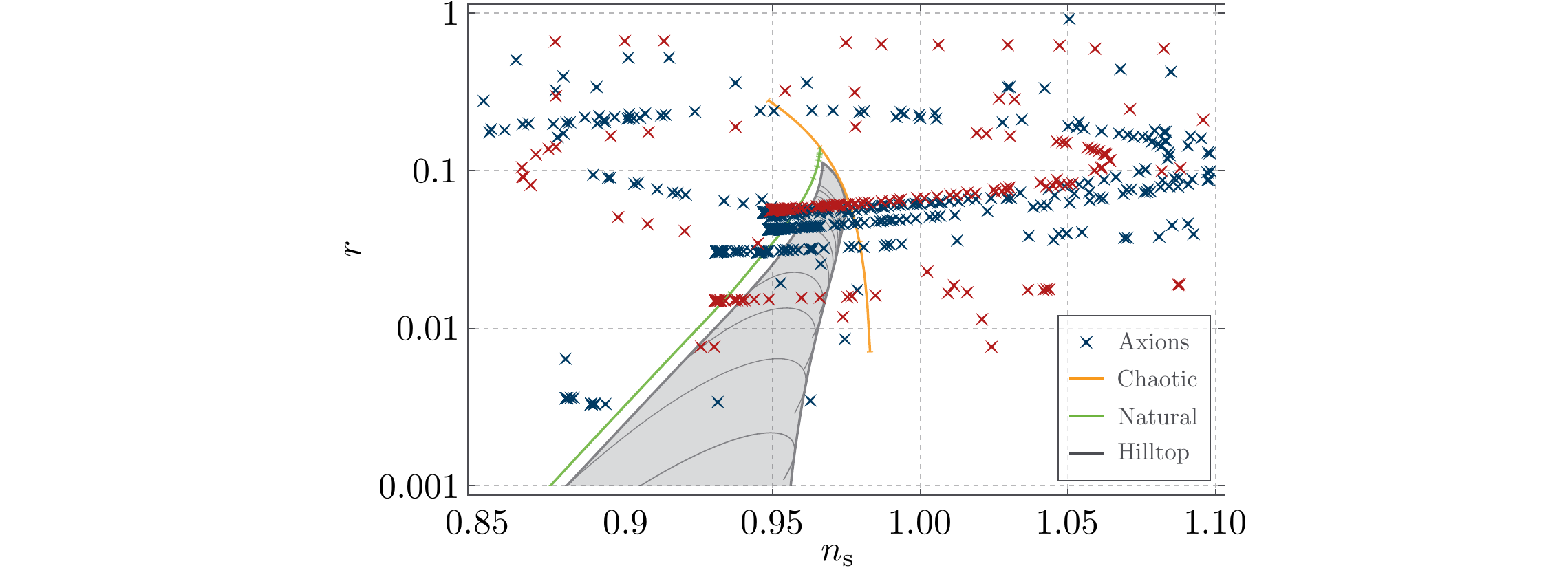}
  \caption{\small Sample of tensor-to-scalar ratios $r$ and spectral indices $n_\text{s}$ for the particular axion theory discussed in \S\ref{infexample}. The red crosses denote trajectories with 60 or more efolds of inflation that terminate in the set of minima with $V_\text{vac} \approx 0.64\Lambda^4$,  while the blue crosses denote $> 60$ efold trajectories that terminate in the vacua with all other vacuum energies in (\ref{vacenergies}). As discussed in the text, for a trajectory terminating in vacuum $i$, $V_0$ is set equal to $-V_{\text{vac},i}$ so that the total energy density vanishes at the endpoint. For comparison we show the observables corresponding to chaotic, natural and hilltop inflation.  Some data points are outside the range displayed.} \label{nsrplane}
\end{figure}
{Let us consider a particularly simple axion potential to illustrate our technique for the systematic sampling of inflationary trajectories. } The theory is discussed at length in the appendix D of \cite{bejk2}, but here we have set the phases to zero.  The relevant parameters are
\be\label{samplethry}
N=2\,,~~P=3\,,~~\bs K \approx \M^{2} \left(\begin{matrix}107 &41.4&21.7\\ 41.4 &48.2 & 47.9\\ 21.7&47.9 & 62.7 \end{matrix}\right)\,,~~ \mathbfcal{Q} =\left(\begin{matrix}1&1\\2&-3\\-3&0\end{matrix}\right)\,,~~\Lambda_I = \Lambda\,,
\ee
where we chose the largest eigenvalue of the kinetic matrix $f_2^2=(11\M)^2$ to allow for inflation, as axion alignment is inefficient at $N=2$.  As discussed above, $V_0$ is chosen to  successively set each of the vacua to vanishing vacuum energy density. We chose initial positions that are uniformly distributed over the periodic domain of the potential, and vanishing initial velocities. Solving the equations of motion (\ref{eominflation}) for the classical trajectories, and selecting all trajectories that terminate at vanishing energy density after more than 60 efolds of inflation, we obtain a representative sample of the inflationary dynamics, as illustrated in Figure \ref{nis3example}. There are nine stable vacua, four of which are doubly degenerate, at vacuum energy densities
\be \label{vacenergies}
\frac{V_{\text{vac}\,,i}}{\Lambda^4} \approx \{ 0, 0.17, 0.64, 1.3, 1.9 \}\,.
\ee
{The field ranges (defined by the distance to the edge of the ``tile" surrounding the minimum, cf.~\cite{bejk2}) along} the lightest direction around each of the vacua are given by 
\bea
{\cal R}_{\text{light},+,i}/\M&=&\{16.4, 17.8, 13.8, 19.6, 18.8\} \,, \nonumber\\
{\cal R}_{\text{light},-,i}/\M&=&\{16.4, 15, 18.9, 11.4, 4.4\} \,.
\eea
These field ranges can be read off from Figure \ref{nis3example}.

Since there are multiple fields active during inflation it is not easy to obtain the correlation functions of perturbations. Still, we can evaluate the spectral index and the tensor-to-scalar ratio assuming single field, slow roll inflation. The resulting observables are shown in Figure \ref{nsrplane}. Clearly a very wide range of observables is possible, even within this extremely simple theory. This is a very direct example of how large the theoretical uncertainties remain, even if we were able to uniquely identify the effective theory governing our landscape.

\begin{figure}
  \centering
  \includegraphics[width=1\textwidth]{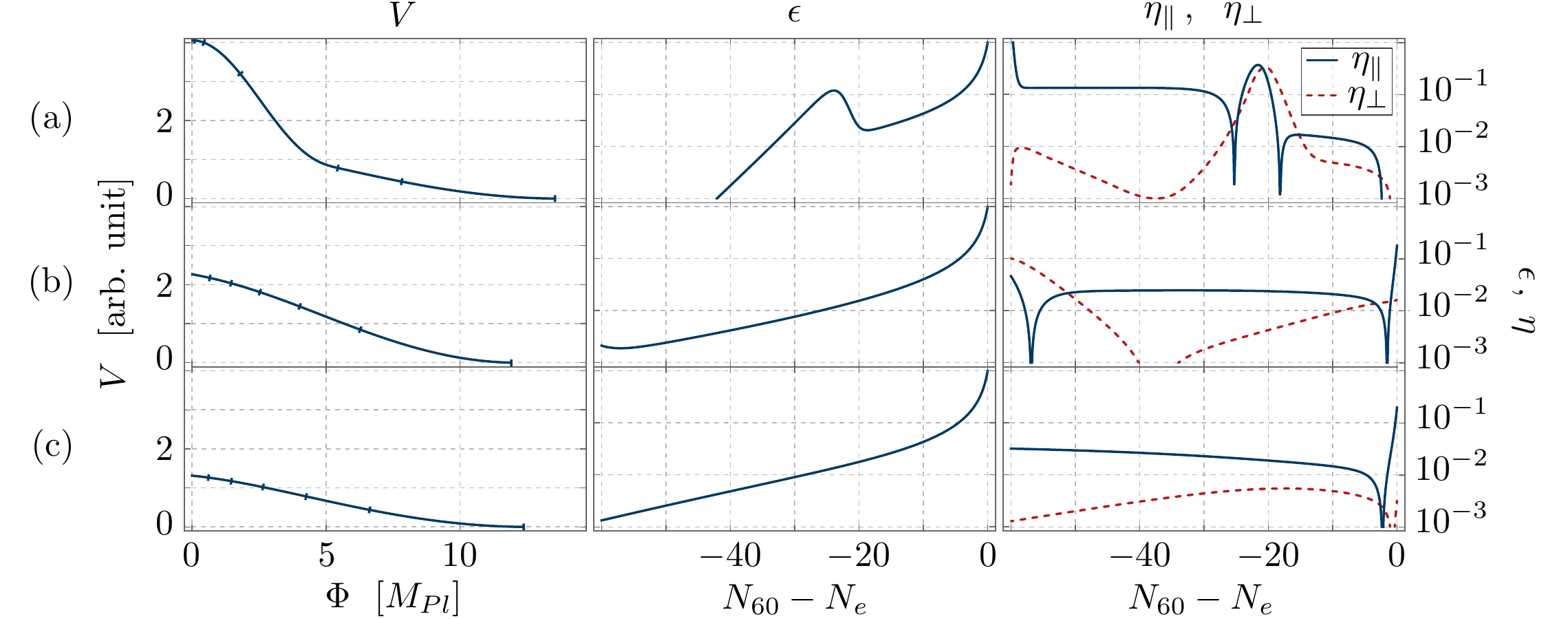}
  \caption{\small Effective potential and slow roll parameters during the last $60$ efolds of three particular trajectories. $N_e-N_{60}$ denotes the number of efolds before the end of inflation. The observable 60 efolds from the end of inflation in the single field, slow roll approximation are: (a) $n_\text{s}=0.73$, $r=5\times 10^{-4}$; (b) $n_\text{s}=0.96$, $r=0.03$; (c) $n_\text{s}=0.93$, $r=0.03$. Ticks along the potential mark $\Delta N_e=10$ intervals. }\label{examples}
\end{figure}
It is important to check whether the single field, slow roll approximation is valid. In Figure \ref{examples} we display the effective potential and the slow roll parameters $\epsilon$, $\eta_\parallel$ and $\eta_\perp$ for the last 60 efolds of inflation of three specific inflationary runs. The examples were chosen to illustrate the wide variety of dynamics: the slow roll parameter $\eta_\perp$ can be large, or small, compared to $\eta_\parallel$. A more sophisticated multifield analysis of the perturbations is required to study the possible non-Gaussianity signatures of these dynamics. Some of the trajectories exhibit turns, and the tensor-to-scalar ratio can vary between $5\times 10^{-4}$ and $1$. Considering the effective potential along the trajectory we see that inflation does not proceed in a quadratic potential. In part this is because we chose the scale of the kinetic matrix such that Planckian displacements are possible, but the quadratic domain alone is not large enough to support $60$ efolds of inflation. We expect this qualitative finding to hold much more generally than in this particular example: inflation in axion theories allows for an extremely wide range of observables and inflationary energy scales. 

\subsection{Inflation after tunneling}\label{inflationaftertunneling}
We now turn to the specific scenario where the inflationary initial conditions originate from the decay of a meta-stable vacuum. Generally speaking, inflation after  barrier tunneling  seems to require fine-tuning.  The condition for thin-wall tunneling is that $M_{\text{Pl}}^2 |V''_*| / V_* \gg 1$ (cf.~(\ref{HMcondition})), while a necessary condition for slow roll is the opposite, that $\eta_V \equiv M_{\text{Pl}}^2 |V''| / V \ll 1$.  These conditions are not logically incompatible because the former applies at the maximum of the barrier $V_*$ while the latter applies to the potential slope after the barrier, but there is nevertheless a clear tension \cite{GarciaBellido:1997uh, Linde:1998iw}.

One of the interesting features of random multi-axion theories is the existence of a hierarchy of Hessian eigenvalues -- the fact that at large $N$ different directions in field space can have very different second derivatives.  Since the least-action path for tunneling tends to coincide with directions in which the barrier is thinnest and the height is lowest, it is plausible that tunneling will proceed in directions where ${|V''_*|}/{V_*}$ is large.  Tunneling in such a direction can leave the field displaced from the minimum along a direction or directions with much smaller ${|V''|}/{V}$ (see also \cite{GarciaBellido:1997dt}).  Therefore inflation after tunneling does not necessarily require tuning beyond the large number of fields $N \gg 1$.  

If the tunneling creates a region with a field value that is in or near the ``quadratic domain" of a low-lying minimum, the  potential will be roughly quadratic.  For a given choice of axion parameters this makes sharp inflationary predictions, as essentially all low-lying minima in the class of theories we are considering are very similar.  For instance, the amplitude of density perturbations will be (\ref{deltarhoquad}).

As discussed in \S \ref{sec:vactransitions}, our analytic control over the potential is strongest for minima that are ``low-lying"; that is, those with vacuum energy less than $\Lambda^4$ above the global minimum of the potential.  When $V_0 \ll \Lambda^4$ this includes all small-CC vacua in which structure can form.  However, it does not include all parent vacua from which the universe might have tunneled to a given small-CC target.  Inflation following tunneling from a high minimum could in principle take place on some feature of the potential outside the quadratic region surrounding the target minimum where the inflationary trajectory should end.  (In \S\ref{samp}  we  analyzed  general inflationary trajectories.)  

We cannot  rule out the possibility that inflationary histories might be dominated by  such non-quadratic potentials, but for the rest of this section we will focus on inflation following tunneling that takes place in the quadratic region. In the quadratic region a necessary condition for at least 60 efolds of inflation is that the typical field range in (\ref{d0estimate}) should satisfy 
\begin{equation}
{\cal R}  \approx   {4 \pi  f\over \sigma_{\mathcal Q}} \frac{ \sqrt{P N}}{P - N} > 15 \M \,.
\end{equation}
Assuming the parameters are such that ${\cal R}$ satisfies this condition, tunneling from high minima should sometimes produce $N_e > 60$ efolds, with a power spectrum set by (\ref{deltarhoquad}).

We can say much more about tunneling between low-lying minima.  Clearly, achieving large amounts of inflation following  a tunneling from a low-lying parent is more difficult, because the starting point on the potential is lower.  Nevertheless we will see that it is possible, albeit with more restrictive conditions on the parameters, and we will exhibit an explicit numerical example.

As discussed in \S \ref{sec:vactransitions}, the dominant tunneling trajectories are generally those between neighboring  minima that are separated by a $2 \pi$ shift in one or a few cosines ($k=1$ or $k=$ few, respectively).  
The mean separation between face neighbor ($k=1$) vacua is\footnote{For a neighbor of degree $k$, the mean vacuum separation scales $\propto\sqrt{k}$, see \S\ref{neighborsec}.}
\begin{equation} \label{PDdistance-sec}
	\left\langle \lVert \bs \Theta_{\text{parent}} - \bs \Theta_{\text{target}} \rVert_2 \right\rangle \equiv \left\langle \lVert \Delta \bs \Theta \rVert_2 \right\rangle \approx \frac{2 \pi f}{\sigma_\mathcal{Q}} \times \mathcal{O}(1) \,.
\end{equation}
Importantly, the distribution of distances has a polynomial tail (see appendix section \ref{appendix-distances}). Vacuum separations significantly larger than   (\ref{PDdistance-sec}) are much more frequent than separations significantly smaller than (\ref{PDdistance-sec}) due to this tail. 

When the separation is greater than $\M$ (as required for inflation in the quadratic regime), we expect roughly half of the vacuum separation to be relevant for a possible period of inflation after the tunneling event. There are two reasons for this.  First, the saddle point is located roughly halfway in between the parent and the target vacuum.  Second,  the field space distance traversed by a Coleman-de Luccia instanton is in general sub-Planckian.  To see this, note that if the instanton enters the regime of slow roll, dimensional analysis suggests that
\begin{equation}
	\Delta \phi = \dot{\phi} \, \Delta \tau \approx \frac{V'}{3H} \Delta \tau \leq \frac{V'}{3H^2} \approx \sqrt{2 \epsilon_V} \, \M \,,\label{eq18}
\end{equation}
where we have used the slow roll equations and the fact that the instanton exists only for a Euclidean time of order $1/H$ (the radius of the four-sphere). If $\Delta \phi \gtrsim \M$, (\ref{eq18}) implies that a slow roll condition would be violated, making a significant support of the instanton in the slow roll regime inconsistent.

 If (as just argued for above) roughly half  the field space separation between the minima is available for inflation following tunneling, the probability to find at least 60 efolds of inflation scales (for $N \gg 1$ and $P-N \ll N$) as
\begin{align} \label{P60}
	\text{Prob}(N_e > 60) & \sim  \text{Prob} \left( \lVert \Delta \bs \Theta \rVert_2 > 30 \M \right)  \notag \\
	&\sim \left( \frac{2 \pi f / \sigma_\mathcal{Q}}{30 \M} \right)^{P-N+1} \,,
\end{align}
where the final scaling is estimated from numerical observations of the tail of the distribution, as discussed in \S \ref{appendix-distances}. 
 
\begin{figure}
  \centering
  \includegraphics[width=1\linewidth]{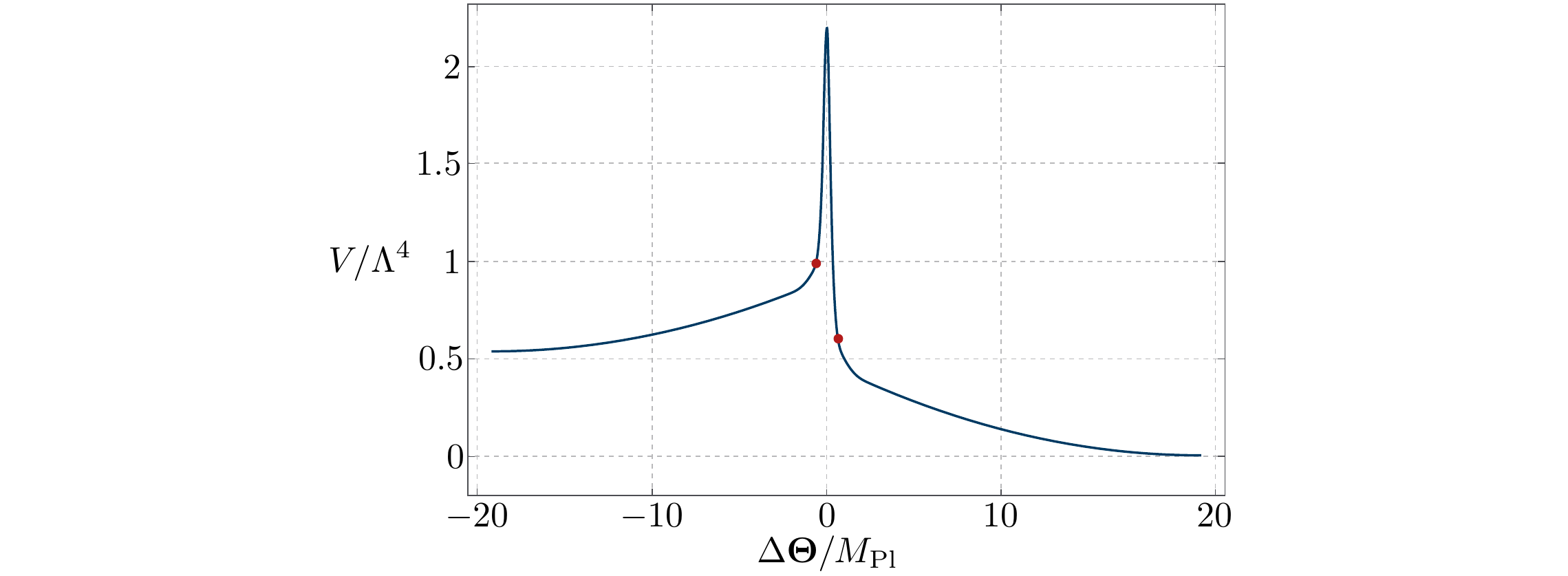}
  \caption{The potential between two ``face neighbor" minima in an example with randomly chosen charges $\mathbfcal{Q}$, plotted along the gradient flow line that passes over the saddle in between. The red dots indicate the two extreme values of the field for the approximate Coleman-de Luccia instanton found numerically using this one-dimensional potential.  There are approximately 109 efolds of inflation following tunneling, and (as expected for inflation post-tunneling) the inflationary parameters 60 efolds from the end are close to those of standard quadratic inflation.  The axion model parameters for this example are $N=100, P=101, \sigma_{\mathcal{Q}}^{2}=1/20, f = 10^{-1} \M$ and $\Lambda = 10^{-2} \M$. Most randomly drawn $\mathbfcal{Q}$ matrices with these parameters will not yield this much inflation after tunneling between face neighbors (cf.~(\ref{P60})); this example was found after ${\cal{O}}(10)$ draws. }
  \label{fig:tunpot}
\end{figure}
We have numerically tested this paradigm by constructing an effective 1D scalar potential which captures the essential properties of the axion potential between the two vacua. Using the  numerical technique described in \S \ref{gradflow}, the 1D effective potential was constructed by following the gradient of the full $N$-dimensional potential starting at the parent minimum and ending at the target minimum, \textit{and} passing directly through the (degree one) saddle in between. This 1D potential indeed has the expected shape:  a super-Planckian range slow roll regime starting at one minimum that connects to a sharp, sub-Planckian range barrier with height $\approx 2 \Lambda^4$, connecting back to another super-Planckian range  slow roll regime that ends at the other minimum. Using this 1D potential we solved the Euclidean Einstein and scalar field equations in the inverted potential to find the instanton, and then solved the Lorentzian Einstein equations in the full $N$-dimensional field space, using the extreme value of the instanton as an initial condition, to find the evolution after tunneling (including any inflation).  We plot an example in Figure \ref{fig:tunpot}.

\subsection{Reheating}\label{reheat}
Axions may interact with gauge fields $F$ via the coupling
\be\label{gaugefieldcoupling}
{\alpha\over8\pi f_{\text{inf}F}} \, \Theta_\text{inf} F\tilde F\,,
\ee
where $\alpha$ is a dimensionless coupling constant and $f_{\text{inf}F}$ is the effective axion decay constant for the inflaton $\Theta_\text{inf}$. The interaction (\ref{gaugefieldcoupling}) is topological when the inflaton evolves slowly, but becomes important at the end of slow roll inflation. This allows an efficient energy transfer from the inflationary to the gauge field sector that can drive reheating. Reheating proceeds through a combination of perturbative and non-perturbative processes. More details will appear in \cite{axidentaluniverse}. 

\section{Light axion phenomenology} \label{lightaxionpheno}
We now turn towards a brief discussion of light axion phenomenology \cite{Hu:2000ke,Arvanitaki:2009fg,Hui:2016ltb}. In particular we will be interested in whether a coupling between QCD and light axions can resolve the strong CP problem and if fuzzy dark matter can be accommodated in a multi-axion theory without fine-tuning.

Thus far we assumed that the leading non-perturbative contributions to the potential (\ref{axion-potential}) stabilize all axions, i.e. the charge matrix $\mathbfcal Q$, or equivalently the canonically normalized charge matrix $\mathbf Q \equiv \mathbfcal Q \bs K^{-1/2}$, is full rank. This is consistent, as very light axions would be frozen by Hubble friction in the early cosmological evolution. In order to study extremely light states, however, let us now assume that the $P$ leading and relatively strong non-perturbative effects stabilize all but $L \ge 1$ of the $N$ axions, that is, $\text{Rank}(\mathbfcal Q) = N - L < N$ (cf. appendix A of \cite{bejk2}). It is generally believed that theories of quantum gravity do not permit global continuous symmetries \cite{Vafa:2005ui,ArkaniHamed:2006dz,Rudelius:2014wla,Cheung:2014vva,Heidenreich:2015wga,Bachlechner:2015qja,Rudelius:2015xta,Hebecker:2015rya,Ibanez:2015fcv,Heidenreich:2016aqi}. Therefore, at least the gravitational axion potential $V_{\text{gr}}$  breaks the $L$ remaining shift symmetries. Since we are interested in light axions, we assume that the energy scale associated to the gravitational potential $V_{\text{gr}}$ is vastly smaller than that of the leading potential, so when the light fields are relevant we can ignore the dynamics of fields stabilized by the $P$ leading non-perturbative effects. The $L$-dimensional subspace relevant for low energy dynamics is then the null space of the leading charge matrix $\mathbf Q$. We define an orthonormal basis $\{ \bs t_1, \bs t_2, \dots , \bs t_L \}$ of $\text{ker}(\mathbf Q)$, 
\be
\mathbf Q \, \bs t_l = \bs 0 \,,~~~\forall \, l \in \{ 1,\dots,L \} \,,
\ee
which we extend to an orthonormal basis of the full field space $\mathbb{R}^N$ by $N-L$ vectors $\bs t_{L+1}, \bs t_{L+2}, \dots, \bs t_N$. We further decompose the field into light and heavy components via
\begin{equation}
	\boldsymbol{\Theta} = \bs T \left( \begin{matrix} \bs \Theta_\text{light} \\ \bs \Theta_\text{heavy} \end{matrix} \right) = \big( \bs T_\text{light} \, \vert \, \bs T_\text{heavy} \big) \left( \begin{matrix} \bs \Theta_\text{light} \\ \bs \Theta_\text{heavy} \end{matrix} \right) \,,
\end{equation}
where we have split $\bs \Theta$ into a piece $\bs \Theta_\text{light}$ of length $L$ and a piece $\bs \Theta_\text{heavy}$ of length $N-L$, and the matrices $\bs T_\text{light,heavy}$ are composed by placing the $\bs t_{1 , \dots, L}$ respectively the $\bs t_{L+1 , \dots, N}$ on consecutive columns, see also \cite{bejk2}.

It will be instructive to consider two distinct sectors that contribute to the subleading non-perturbative potential: the axions couple to QCD instantons, as well as gravitational wormhole instantons \cite{Giddings:1989bq}. The former coupling gives rise to a Peccei-Quinn (PQ) axion that can solve the strong CP problem \cite{Peccei:1977hh}, while the latter coupling is motivated by popular beliefs about quantum gravity. As pointed out in \cite{Hui:2016ltb}, gravitational strength breaking of axion shift symmetries might result in a viable candidate for ultra-light axion dark matter. However, the two mechanisms might seem mutually exclusive. The subleading contributions couple to all axions, including the QCD axion and there is no reason to expect that the relative phase between gravitational and gauge instantons is correlated. This imposes an upper bound on the strength of gravitational instantons in order not to spoil the PQ solution to the strong CP problem \cite{Svrcek:2006yi}. On the other hand, the QCD axion could potentially give rise to an overwhelming amount of phenomenologically unacceptable dark matter, which in turn strongly constrains the effective decay constant or the initial misalignment of the axion acting as a PQ symmetry (see e.g. \cite{Marsh:2015xka}).

\subsection{Light and lightest axions}
We write the potential relevant for the $L$ canonically normalized light axions $\bs\Theta_\text{light}$ as
\begin{align} \label{slpotential}
V_{\text{gr}}(\bs\Theta_\text{light}) = \Lambda_{\text{QCD}}^4 &\left[ 1 - \cos \left( \mathbf Q_\text{QCD} \bs T_\text{light} \bs \Theta_{\text{light}} + \delta_{\text{QCD}} \right) \right] + \notag \\ &\sum_\alpha \M^4 \, e^{-S^\alpha} \left[ 1 - \cos\left( \mathbf Q_\text{gr} \bs T_\text{light} \bs\Theta_{\text{light}} + \bs\delta_{\text{gr}} \right)^\alpha \right] \,,
\end{align}
where $\alpha$ is an index running over the subleading contributions to the axion potential and $ \mathbfcal Q_\text{gr}$ and $ \mathbfcal Q_\text{QCD}$ are the integer charges of the gravitational and QCD instantons, respectively. We've also defined the canonical charge matrices
\be
\mathbf Q_\text{gr} = \mathbfcal Q_\text{gr} \bs K^{-1/2} \,, ~~ \mathbf Q_\text{QCD} = \mathbfcal Q_\text{QCD} \bs K^{-1/2} \,,
\ee
where ${\mathbfcal Q}_{\text{gr}}$ is the integer charge matrix of the gravitational instantons. In defining the coordinates $\bs \theta$ in (\ref{axionlagrangian}) we assumed that any instanton has integer charges with respect to all axions. Gravity indiscriminately couples to all axions, so we expect that ${\mathbfcal Q}_{\text{gr}}$ contains the entire integer lattice $\mathbb Z^N$. Since the instanton actions depend on the instanton charges, for large instanton actions only a small number of gravitational instantons will provide a relevant contribution. Let us consider instantons whose action can be parametrized as
\be\label{slinstantonaction}
S^\alpha = {\mathcal S} \M \lVert \mathbf Q_\text{gr}^\alpha \rVert_2 + \delta S^\alpha = {\mathcal S} \M \sqrt{\mathbf Q_\text{gr}^\alpha \, (\mathbf Q_\text{gr}^\alpha)^\top} + \dots \,,
\ee
where the ellipses denote corrections to the classical instanton action that we will ignore and, for clarity, $\mathbf Q_\text{gr}^\alpha = (\mathbf Q_\text{gr})^\alpha_{~i=1,\dots,N}$ is the $\alpha^\text{th}$ row of $\mathbf Q_\text{gr}$ which contains the couplings of the $\alpha^\text{th}$ gravitational instanton to the $N$ axions in the $\bs \Theta$-basis. For the example of Euclidean wormholes the prefactor of the classical action is given by ${\mathcal S}=\sqrt{6}\pi/8\approx 1$ \cite{ArkaniHamed:2007js,Madrid,Bachlechner:2015qja}. We note that while we can compute the action for Euclidean wormholes, our argument holds more generally as long as the parametrization (\ref{slinstantonaction}) is valid and ${\mathcal S}\approx {\mathcal O}(1)$.

While it is generally believed that quantum gravity breaks all continuous global symmetries, it is not immediately obvious to what extent these symmetries are broken. One natural guess is to assume that all gauge interactions mediate forces that are no weaker than the gravitational interactions. This assumption is known as the weak gravity conjecture \cite{ArkaniHamed:2006dz}. The conjecture has been generalized to multi-axion theories in \cite{Cheung:2014vva}, where it becomes the requirement that the convex hull of the vectors
\be
\bs z^a = { \mathbf Q^a \over S^a } \M \,,
\ee
contains the unit ball.\footnote{The index $a$ runs over \textit{all} the non-perturbative contributions to the potential, and $\mathbf Q$ here is the full canonically normalized charge matrix (i.e. it contains the rows of what we have previously called $\mathbf Q$, namely the canonically normalized charge matrix of leading instantons, $\mathbf Q_\text{gr}$ and $\mathbf Q_\text{QCD}$).} Note that for the actions (\ref{slinstantonaction}) all vectors $\bs z^a$ have length $1/{\mathcal S}$, so that whenever ${\mathcal S}<1$ the convex hull condition of the weak gravity conjecture is satisfied. Perhaps not surprisingly, gravitational instantons roughly saturate this bound.

Consider now the case of $L=2$ light axions. Whenever the QCD instantons constitute an important contribution to the potential of the light axions we can simply write the relevant potential as
\be
V_{\text{gr}}(\tilde{\bs \Theta}_\text{light})=\Lambda_{\text{QCD}}^4\left[1-\cos\left( {\Theta_\text{QCD}\over f_{\text{QCD}}}+\delta_{\text{QCD}} \right) \right]+\M^4 \, e^{-S^1}\left[1-\cos\left({\Theta_\text{DM}\over f_{\text{DM}}}+\delta_\text{gr}^1 \right) \right]\,,
\ee
where $S^1$ denotes the action corresponding to the most important gravitational instanton contribution and we defined the fields $\tilde{\bs \Theta}_\text{light} = (\Theta_\text{QCD}, \Theta_\text{DM})$ by the linear combinations
\be
{\Theta_\text{QCD}\over f_{\text{QCD}}}\equiv \mathbf Q_\text{QCD} \bs T_\text{light} \bs\Theta_{\text{light}}\,,~~~{\Theta_\text{DM}\over f_{\text{DM}}} \equiv \mathbf Q^1_{\text{gr}} \bs T_\text{light} \bs\Theta_{\text{light}}\,,
\ee
and we assumed $\Lambda_{\text{QCD}}^4 \gg \M^4 \, e^{-S^1}$ so we can drop all other gravitational contributions. The relevant axion decay constants are given by
\be\label{lightaxionfs}
f_{\text{QCD}}=\lVert \mathbf Q_\text{QCD} \bs T_\text{light} \rVert_2^{-1}\,,~~f_{\text{DM}}=\lVert \mathbf Q^1_\text{gr} \bs T_\text{light} \rVert_2^{-1}\,.
\ee
Note that the transformations leading to the light axions $\Theta_\text{QCD}$ and $\Theta_\text{DM}$ are not orthogonal, so kinetic couplings can remain to the heavy fields. However, in the parameter regime we are interested in there is a vast hierarchy in the axion masses, such that the heavy fields will be stationary when the light fields are dynamical and we can ignore these kinetic couplings.

\subsection{Fuzzy dark matter}
Let us discuss the impact of the lightest axion $\Theta_\text{DM}$ on the cosmological evolution. This axion is frozen during most of the cosmological history, but begins to oscillate when the Hubble scale drops below the axion mass. After this time, the axion will oscillate in its approximately quadratic potential and act as dark matter. In this section we merely comment on some generic features, and defer a more detailed discussion to \cite{axidentaluniverse}.

In principle, the gravitational sector contributes an infinite number of terms to the non-perturbative potential: the charges ${\mathbfcal Q}_{\text{gr}}$ contain all sites of the lattice $\mathbb Z^N$. However, in the regime of perturbative control only a small number of those terms will be relevant for low-energy physics. Given the instanton actions (\ref{slinstantonaction}), the leading non-perturbative effects correspond to those sites in the charge lattice with smallest two-length $\lVert \mathbf Q_\text{gr}^\alpha \rVert_2$. In the simple ensemble where $\bs K = f^2 \mathbbold{1}$,\footnote{Our results continue to hold qualitatively if a less trivial ensemble of kinetic matrices is considered, for example the Wishart ensemble, cf. \cite{Bachlechner:2015qja}.}
\begin{equation} \label{minkbound}
	\underset{\alpha}{\text{min}} \left( \lVert \mathbf Q_\text{gr}^\alpha \rVert_2 \right) = 1/f \,.
\end{equation}
With this we can estimate the most important contribution to the lightest axion $\Theta_\text{DM}$. To that end, let us approximate the light field space directions $\bs T_\text{light}$ as isotropic and independent of the charges corresponding to the most important gravitational instanton. In general this approximation is violated: both $\bs T_\text{light}$ and $\mathbf Q_\text{gr}$ depend on the kinetic matrix $\bs K$, and therefore can be correlated to some extent. Assuming this correlation can be neglected, we expect $f_\text{DM} \approx \sqrt{N/2} / \lVert \mathbf Q_\text{gr}^1 \rVert_2$. With (\ref{slinstantonaction}), (\ref{lightaxionfs}) and (\ref{minkbound}) we then have for the dark matter decay constant and corresponding instanton action at large $N$,
\be \label{dmaction}
f_{\text{DM}} \approx \sqrt{\frac{N}{2}} f \,,~~S^1\approx {\cal S} {\M\over f}\,.
\ee
The mass of the light axion is given by
\be
m_{\text{DM}}={\M^2\over f_\text{DM}} e^{-S^1/2}\,.
\ee

It may be instructive to consider the case of $N \approx 200$ axions and a GUT scale decay constant $f \approx 4 \times 10^{-3} \M$. These values give a light axion mass $m_{\text{DM}} \approx 10^{-22}\,\text{eV}$, consistent with the mass required to realize fuzzy dark matter. The expected axion decay constant $f_{\text{DM}} \approx 4 \times 10^{-2} \M$ coincides with the value giving the desired dark matter abundance  \cite{Arvanitaki:2009fg,Hui:2016ltb},
\be
\Omega_{\text{axion}} \sim 0.2 \left({ f_{\text{DM}} \over .04  \,\M}\right)^{2}  \left({m_{\text{DM}} \over 10^{-22} \, \text{eV}} \right)^{1/2}\,.
\ee

\subsection{QCD axion}
In order to estimate the decay constant relevant to the QCD axion, we again first estimate the two-norm $\lVert \mathbf Q_\text{QCD}\rVert_2$. We have, in the $\bs K = f^2 \mathbbold{1}$ ensemble,
\be
\lVert \mathbf Q_\text{QCD}\rVert_2 = \lVert {\mathbfcal Q}_{\text{QCD}} \bs K^{-1/2} \rVert_2 = {{\sigma_\text{QCD}} \over f} \,,
\ee
where ${{\sigma_\text{QCD}}}$ denotes the root-mean-squared value of the entries in ${\mathbfcal Q}_{\text{QCD}}$. Assuming again that the light directions $ \bs T_\text{light} $ are uncorrelated with the canonically normalized charges of the QCD axion we roughly expect $\lVert \mathbf Q_\text{QCD} \bs T_\text{light} \rVert_2 \sim \lVert \mathbf Q_\text{QCD} \rVert_2 / \sqrt{N/2}$, or equivalently
\be
f_{\text{QCD}}\sim \sqrt{\frac{N}{2}} \frac{f}{\sigma_\text{QCD}} \,.
\ee
Finally, the non-perturbative contribution due to QCD stabilizes the corresponding axion at $-\delta_{\text{QCD}}f_{\text{QCD}}$, eliminating the CP-violating phase. However, there exist a large number of gravitational instantons that also couple to the same fields and potentially introduce a large amount of CP violation. Very roughly, we can estimate the shift of the phase due to the gravitational instantons as
\be
{\Delta\Theta_{\text{QCD}}\over f_{\text{QCD}}}\lesssim {\M^4 e^{-S^1}\over \Lambda_{\text{QCD}}^4}\,.
\ee
Using $\Lambda_{\text{QCD}}^4 \approx 10^{-78}\M^4$ and the estimate (\ref{dmaction}) we have
\be
\log_{10}\left({\Delta\Theta_{\text{QCD}}\over f_{\text{QCD}}}\right)\sim 78 - \frac{{\cal S}}{\log 10} {\M\over f}\,.
\ee
Using the values from above, we see that the gravitational instantons contribute a CP-violating phase of roughly
\be
{\Delta\Theta_{\text{QCD}}\over f_{\text{QCD}}}\sim 10^{-24} \ll 10^{-10}\,,
\ee
where the last inequality denotes the comparison to the experimental bound. Hence gravitational instantons do not spoil the PQ solution to the strong CP problem.

Note that the QCD axion might also act as dark matter and potentially lead to an overwhelming amount of phenomenologically unacceptable dark matter, depending on the initial misalignment angle \cite{Marsh:2015xka}. The severity of this tuning  depends on the parameter choice, and any accidental alignment between the axions. We will return to this issue in the more comprehensive discussion in \cite{axidentaluniverse}.

\section*{Acknowledgements}
We thank Mafalda Dias, Jonathan Frazer, Ben Freivogel, Mark Hertzberg, Liam McAllister, Marjorie Schillo, Alex Vilenkin and Jeremy Wachter for useful discussions at various stages of the development of this paper. The work of TB is supported in part by DOE under grants no. DE-SC0011941 and DE-SC0009919 and by the Simons Foundation SFARI 560536. The work of KE is in part supported by the INFN. OJ is supported by a James Arthur Graduate Fellowship. The work of MK is supported in part by the NSF through grants PHY-1214302 and PHY-1820814. This work was performed in part at the Aspen Center for Physics, which is supported by National Science Foundation grant PHY-1066293.

\appendix
\section{Neighbors and saddle points}\label{appendix-saddles}
In this appendix we provide a derivation of two results used in \S \ref{sec:vactransitions} and \S \ref{inflationaftertunneling}. 

The first  result is  the large $N$ behavior of the Hessian eigenvalues of degree-$k$ saddle points of the potential.  These eigenvalues are important because they determine the sharpness of the barrier separating adjacent minima, and hence the characteristics of the instanton that mediates the decay.   This is particularly important to determine the lifetime of a low-CC minimum (so that the ``parent" minimum has nearly zero CC, and the ``target" has lower, probably negative, vacuum energy).  Because of the large number of possible decay channels  the \emph{distribution} of these eigenvalues is important, not just the mean. 

The second result is the large $N$ behavior of the canonically normalized field space distance between two face neighboring, low-lying vacua.  This is important for the question of slow roll inflation following tunneling between two such neighbors (although we emphasize again that decay from a higher minimum or other point on the potential might give rise to inflation even when decays between face neighbors do not).  Of particular importance is the fact that the distribution of the distances between face neighbors is heavy tailed, so that the extreme cases can be much larger than the mean.  For this question the target minimum is the one with nearly zero CC, while the parent has larger, positive CC (although  small enough for the quadratic approximation to be valid).

To avoid confusion we use  subscripts A and B for the vacua, only specifying which has higher energy when necessary. Familiarity with the geometric picture developed in \cite{bejk2}, where the field space is identified with an $N$-dimensional hyperplane $\Sigma$ slicing through a uniform lattice in $P$-dimensional Euclidean space, is assumed.

This appendix is organized as follows.  In \S \ref{degk} we review the necessary technology from \cite{bejk2}, in \S \ref{appendix-hessian} we study the distribution of Hessian eigenvalues in detail, and in \S \ref{appendix-distances} we study the distribution of face neighbor distances.

\subsection{Degree-$k$ saddle points}\label{degk}
To study tunneling to/from a quadratic domain vacuum $\bs \Theta_{\text{A}}$ in a tile labeled by $\bs n_{\text{A}}$, we first identify its nearby tiles. We expect that the dominant decay channels will be between $\bs \Theta_{\text{A}}$ and its neighboring vacua. What constitutes a ``neighboring" vacuum is to some degree arbitrary, but 
a reasonable definition adopted in  \cite{bejk2} is to deem vacua  in tiles whose $P$-cubes are separated by 
at most one step in each of the $\phi^J$-coordinate directions as ``neighbors", i.e. those at
 \begin{equation}\label{all-neighbors}
2\pi \bs{n}_{\text{B}}=2\pi\bs{n}_{\text{A}}+2\pi\bs v \,,
\end{equation}
where $\bs v$ is any $P$-vector with entries in $\{-1,0,1\}$, and  $\bs n_{\text{A}}$ and $\bs n_{\text{B}}$ are the integer $P$-vectors for the vacuum and neighbor's lattice points, respectively. There are a total of $3^P$  candidates tiles, but only the subset of $P$-cubes  that intersect $\Sigma$ may actually correspond to  neighboring vacua of $\bs \Theta_{\text{A}}$. 

We can label the  displacement vectors $2\pi \bs v$ by the number of non-zero entries they contain, i.e. the number of mutually orthogonal steps taken from $2\pi \bs n_\text{A}$ in the ambient space to reach the center of the neighbor cube. We indicate this by a natural number subscript $k$, writing $\bs v_k$ from now on. For a given  $k$ there are a total of ${2^k}{ {P}\choose{k}}$ distinct vectors $\bs v_k$. A degree-$k$ saddle point of $V_{\text{aux}}$\footnote{The auxiliary potential is defined as the following function of $P$ scalars,
\begin{equation}
	V(\bs \phi)\equiv\sum_{I=1}^{P}\Lambda_I^4\left(1-\cos(\phi^I)\right) \,.
\end{equation}
The axion potential can be identified as the auxiliary potential evaluated at $\bs \phi=\mathbfcal{Q}\bs \theta$. See \cite{bejk2} for further details.} lies exactly half-way between the pair of auxiliary minima $2\pi \bs n_{\text{A}}$ and $2\pi (\bs n_{\text{A}}+\bs v_k)$, at
\begin{equation}\label{saddle-locations}
\pi \bs n_{\text{S,k}}\equiv2\pi \bs n_{\text{A}}+\pi \bs v_k \,.
\end{equation}
This is because $\bs n_{\text{S},k}$ contains $P-k$ even integers and $k$ odd integers, resulting in $k$ of the cosine terms of $V_{\text{aux}}$ being maximized at $\bs \phi=\pi \bs{n}_{\text{S},k}$, while the rest are minimized. In the vicinity of the auxiliary saddle  all $P$ cosines are well-approximated by quadratics. In other words, there are quadratic domains situated not only at the points of the auxiliary lattice $2\pi \mathbb{Z}^P$ (the auxiliary minima), but more generally at the points $\pi \mathbb{Z}^P $ (auxiliary critical points).

For very low values of $k$, \emph{any} of the $2^k{{P}\choose{k}}$ integer displacement vectors $\bs v_k$ will typically identify a nonempty neighboring tile of $\bs \Theta_\text{A}$. This follows from the fact that $\Sigma$ is randomly oriented with respect to the standard basis elements $\{\bs e^{(J)}\}$ of $\mathbb{R}^P$, and so the size of $\Po^\perp \bs e^{(J)}$ is $\sim\sqrt{\nu/P}\ll 1$, for all $J$.\footnote{Note that this is a much weaker notion of alignment than that of the $\bs t^\nparallel_a$ of the aligned lattice basis, which have \emph{exponentially} small perpendicular components.} The expected length of $\bs  v_k^\perp \equiv \Pob \bs v_k$ is 
\begin{equation} \label{vp}
\Vert \bs v_k^\perp \Vert_2 \equiv \Vert \Pob \bs v_k \Vert_2 = \lVert \sum_{J=1}^P  v_k^J  \Pob \bs e^{(J)} \lVert_2 \approx \sqrt{k \nu/P} \,,
\end{equation}
where the $\approx$ follows since this is a sum of $k$ non-zero vectors $\Pob \bs e^{(J)}$ of random orientation and length $\sim\sqrt{\nu/P}$.

 The largest distance one can move away from $\Sigma$  by shifting   from lattice point $2\pi \bs n_{\text{A}}$ by $2\pi \bs v_k$ is $2\pi \lVert  \bs v_k^\perp \rVert_2$. This is just a statement of the triangle inequality: the two-norm distance of a candidate lattice point is bounded by
\begin{equation}
d_{\text{B}}\lesssim d_{\text{A}} +2\pi\sqrt{k\nu/P} \,, \label{neighbor-dist-bound}
\end{equation}
where $d_{\text{A}}$ is the $\ell_2$-distance between the original vacuum in $\Sigma$ and its lattice point in $\mathbb R^P$,
\begin{equation}
d_{\text{A}}\equiv \lVert \bold Q \bs\Theta_{\text{A}}-2\pi \bs n_{\text{A}} \rVert_2 \,.
\end{equation}
Since the vacuum $\bs \Theta_{\text{A}}$ admits a quadratic description, $d_{\text{A}}$ is small (the vacuum energy in the $V_0=0$ theory is proportional to the two-norm distance in the case of equal couplings).  

Concretely,
\begin{equation}
	\frac{1}{2}d_\text{A}^2=2\pi^2 \lVert \Pob \bs n_\text{A} \rVert_2^2< \mu_{\text{max}} P \,, ~~ \mu_{\text{max}}\ll 1 \,.\footnote{$\mu_{\text{max}}$ is defined as the maximum quadratic domain vacuum energy, divided by the average height of the potential, here $P\Lambda^4$. For example, $\mu_{\text{max}}\equiv \frac{V_{\text{max}}}{P\Lambda^4}$ resulted in a value of about $\mu_{\text{max}}=0.014$ in an $N=100$ and $\nu=1$ theory \cite{bejk2}. The value is determined by the parameters and one's choice for the threshold value on the $\ell_{\infty}$-displacement in their definition of the quadratic domain. The quantity $\mu\equiv \frac{V_{\text{vac}}}{P \Lambda^4}$ is useful in that it enables a comparison between quadratic domain vacua of different landscapes by removing the routine factor of $P$ arising simply from the total number of cosine terms, which may differ in the two landscapes.}
\end{equation}
So, for $k\nu/P\ll 1$ we can conclude $d_{\text{B}}$ remains small for all the neighbor candidates defined by (\ref{all-neighbors}).

As $k$ increases the bound on $d_{\text{B}}$ from the triangle inequality weakens, but for any $k$ one can expect some fraction of the  ${2^k}{ {P}\choose{k}}$ candidates to remain close to $\Sigma$. For such $\bs v_{k}$ (those which either have particularly short $\bs v_k^\perp$ themselves, or those whose $\bs v_k^\perp$ are largely canceled by $\Pob \bs n_{\text{A}}$) it certainly follows that the auxiliary saddle between the neighboring lattice points is \emph{also} in close proximity to $\Sigma$. The reason is because the auxiliary saddle's distance from $\Sigma$ is bounded above and below by $d_{\text{A}}$ and $d_{\text{B}}$, respectively, or vice-versa.

In other words, for general $k$ not all neighboring $P$-cubes defined by (\ref{all-neighbors}) necessarily give rise to quadratic domain vacuum neighbors, but the fraction that \emph{do} are separated from $2\pi \bs n_{\text{A}}$ by a degree-$k$ saddle of the auxiliary potential which itself has a quadratic domain that is intersected by $\Sigma$. Consequently, the physical potential in the region between the $\bs n_\text{A}$ and $\bs n_\text{B}$ tiles is well-described by the orthogonal projection of the specific auxiliary saddle domain. Thus, the physical potential has a degree-$k$ saddle point with ambient coordinates $\bold Q\bs\Theta_{\text{S},k}\approx\pi \bs n_{\text{S},k}$, or in canonical coordinates, 
\bea
\bs \Theta_{\text{B}}& \approx  \bs \Theta_{\text{A}}+ 2\pi  (\bold Q^\top \bold Q)^{-1} \bold Q^\top {\bs v}_k\label{parent-location2}\\
\bs \Theta_{\text{S},k}&\approx \bs \Theta_{\text{A}}+\pi (\bold Q^\top\bold Q)^{-1}\bold{Q}^\top {\bs v}_k \,. \label{saddle-loc}
\eea
where we've used $\bold  Q\bs\Theta_{\text{A}}\approx2\pi \bs n_{\text{A}}$.

Furthermore, the Hessian of the physical potential $V$ evaluated at (physical) saddle points $\bs \Theta_{\text{S,k}}$ can be approximated in a simple manner \cite{bejk2}. This was used in \S \ref{sec:vactransitions} in estimating the stability of quadratic domain vacua $\bs \Theta_{\text{A}}$. First, note the chain rule implies a simple expression for the Hessian of the physical potential $H_{ij}\equiv\partial_i\partial_j V$ in terms of the auxiliary one, $H_{IJ}\equiv\partial_I\partial_J V_{\text{aux}}$. In canonical coordinates,
\begin{equation}
\bs{H}_{\text{phys}}(\bs \Theta)=\bold Q^\top \bs{H}_{\text{aux}}|_{\bs \phi=\bold Q \bs{\Theta}} ~\bold Q \,,
\end{equation}
where 
\begin{equation}
\bs H_{\text{aux}}(\bs \phi) = \Lambda^4 \, \text{\diag}\{\cos(\phi^1),\dots, \cos(\phi^P)\} \,.
\end{equation}
At the saddle points (\ref{saddle-locations}), where 
\begin{equation}
\bold Q\bs\Theta_{\text{S},k}\approx 2\pi \bs n_\text{A}+\pi\bs v_k \,,
\end{equation}
we therefore have
\begin{equation}
\bs H_{\text{phys}}(\bs\Theta_{\text{S},k})\approx \Lambda^4 \bold Q^\top \bs D\bold Q \,,
\end{equation}
where $\bs D$ is a diagonal matrix containing $k$ negative ones and $P-k$ positive ones along the diagonal because
\begin{equation}
 \bs H_{\text{aux}}(\bold{Q}\bs \Theta_{\text{S},k})\approx \bs H_{\text{aux}}(2\pi \bs n_\text{A}+\pi\bs v_k)=\Lambda^4\bs D \,.
\end{equation}
The most negative Hessian eigenvalue at saddle points of $V$ is important in our estimate of the CdL decay rate. Our analytical tools are only applicable to physical saddles well-described by the orthogonal projection of auxiliary ones. As noted, this will in general include higher values of $k$ as well. So, we proceed by studying the eigenvalue distributions of ensembles of matrices $\bold Q^\top \bs D\bold Q$ for the full range of $k$, from $1$ to $P$.

\subsection{Hessian eigenvalues at large $N$} \label{appendix-hessian}
It is conducive to pull out the overall scales introduced trivially by the random charge matrix $\mathbfcal{Q}$ and kinetic matrix $\bs K $. To that end, define $\hat{\mathbfcal{Q}}$ such that
\begin{equation}
\mathbfcal{Q}=\sigma_{\mathcal{Q}}\hat{\mathbfcal{Q}} \,.
\end{equation}
Then $\bold Q$, the charge matrix in canonical coordinates, is expressed in terms of a Wigner matrix $\hat{\mathbfcal{Q}}$ whose entries are normally distributed with variance $1$, as follows
\begin{equation}
\bold Q=\frac{\sigma_{\mathcal{Q}}}{f}\hat{\mathbfcal{Q}} \,.
\end{equation}
Then the (physical) Hessian is
\begin{equation}
\bs H=\Lambda^4\left(\frac{\sigma_{\mathcal{Q}}}{f}\right)^2\hat{\bs H} \,,
\end{equation}
where
\begin{equation}
\hat{\bs H}\equiv \hat{\mathbfcal{Q}}^\top \bs D\hat{\mathbfcal{Q}} \,. \label{eq-H-hat}
\end{equation}
Thus it suffices to study the ensemble (\ref{eq-H-hat}) generated by $\hat{\mathcal{Q}}^i_{\,J} \sim \mathcal{N}(0,1)$ at large $N$ (provided $\sigma_\mathcal{Q}$ is not too small, $\gtrsim \sqrt{3/N}$ \cite{Bachlechner:2014gfa,bejk2}).

The spectrum of $\hat{\bs{H}}$ for $k<P$ negative signs in $\bs D$ must in some sense be bounded by a pair of Marchenko-Pastur distributions; one located on the positive  axis and one on the negative axis. This is simply because  in the two extremes, $k=0$ or $k=P$, the matrix $\bs D$ is either plus or minus the identity, making $\hat{\bs H}$ exactly plus or minus the Wishart matrix $\hat{\mathbfcal{Q}}^\top\hat{\mathbfcal{Q}}$.

In fact, it suffices to study just the evolution of these distributions over half of the $k$ values. For even $P$, the range $k=0$ to $k=P/2$ suffices, and for odd $P$, $k=0$ to $k=\frac{P-1}{2}$ does, because $\bs D_{P-k}=-\bs D_{k}$. This means the series of $\hat{\bs H}$ eigenvalue distributions for the second half of $k$ values is given by reflecting each of the distributions in the first half-series onto the negative axis (and then reordering the series of distributions in reverse).

To simplify discussion, let us take $P$ to be even. The numerics confirm the naive expectation of a gradual transition between the Marchenko-Pastur distribution with support on the interval $\sim[1/N,\alpha N]$ for $k=0$, to the -- necessarily symmetric -- distribution for $k=P/2$, whilst at each $k$ value always keeping the (possibly disjoint) support interval contained within $[-\alpha N,\alpha N]$. The Marchenko-Pastur distribution for the  Wishart ensemble (for all fixed $\nu = P- N$) has a right-edge at $\alpha = 4$, and this holds approximately at finite but large  $N$.

\begin{figure}
  \centering
  \includegraphics[width=1\linewidth]{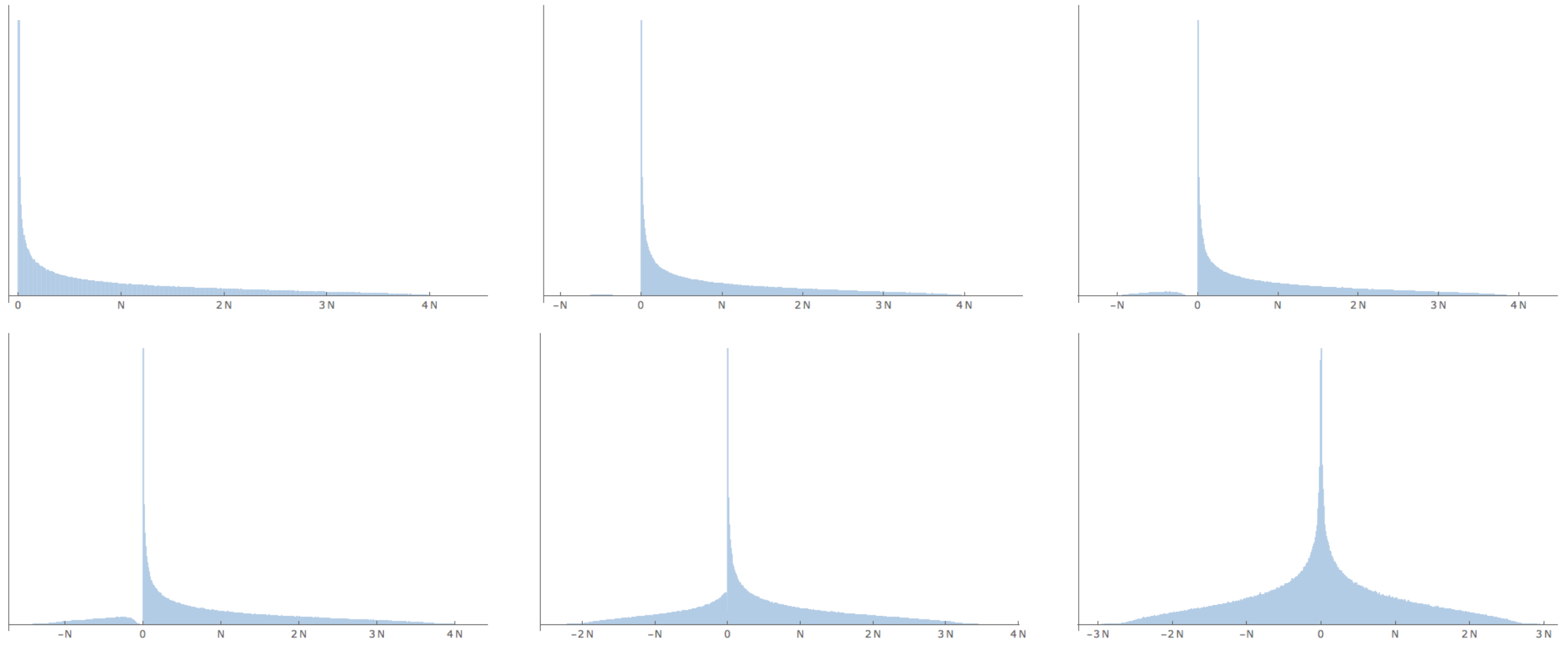}
  \caption{Spectrum of the ``sign-flipped Wishart ensemble" defined by $\hat{\mathbfcal{Q}}^\top \bs D\hat{\mathbfcal{Q}}$ where $\mathbfcal{Q}$ is an i.i.d.~$P \times N$ matrix with real, normally distributed entries and $\bs D$ is a $P \times P$ diagonal matrix with entries $\pm 1$, as the number of negative signs in $\bs D$ increases from $k=1$ to $k=P/2$ from the upper left to lower right.}
  \label{fig:H-eigs-flips}
\end{figure}

The evolution in  $\hat{\bs H}$ eigenvalue distributions is shown in Figure \ref{fig:H-eigs-flips}. Each time $k$ increases the shape of the PDF changes as follows: a layer of the once-Marchenko-Pastur distribution on the positive axis (corresponding to $k=0$) is shaved off in a uniform fashion, and its area is deposited onto the negative axis, beginning with a small bump at $-N/2$ ($k=1$), which ultimately grows into the reflected Marchenko-Pastur shape (on the negative axis). The symmetric distribution that results from flipping exactly half the signs in $\bs D$ is shown in the bottom-right panel of Figure \ref{fig:H-eigs-flips}. 

Turning to the behavior of the minimum eigenvalue of $\hat{\bs H}$, which we'll denote by $\lambda_-$, consider again the extreme case $k=P$. Here the set of most negative eigenvalues of an ensemble of $\hat{\bs H}$ is given by the set of \emph{maximum} eigenvalues of the Wishart ensemble $\hat{\mathbfcal Q}^\top\hat{\mathbfcal{Q}}$, just multiplied by negative one. For $N \gg 1$ and $P-N \ll N$ we have
\begin{equation}
	\langle\lambda_{-}\rangle \approx - \alpha N \,,
\end{equation}
with $\alpha \approx 4$, with standard deviation of order $N^{1/3}$ (e.g. \cite{2007JPhA...40.4317V}). 

\begin{figure}
  \centering
 \includegraphics[width=1\linewidth]{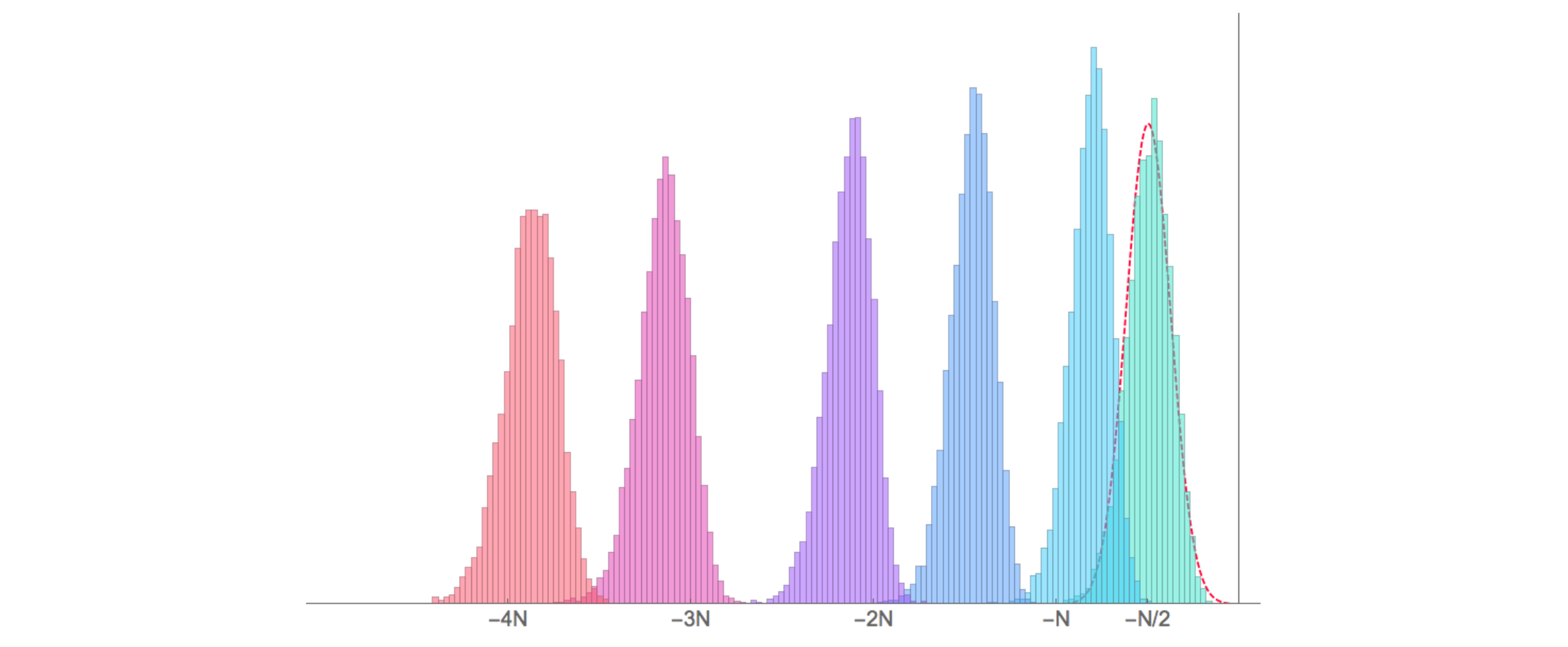}
  \caption{Minimum eigenvalue distributions of $\hat{\bs H}$ ensembles defined as $\hat{\mathbfcal{Q}}^\top\bs D \hat{\mathbfcal{Q}}$ for $N=100$ and $\nu=3$. Moving from right to left are results for $k=1,\thinspace 5, \thinspace 20,\thinspace 40,\thinspace 75\text{, and }100$ negative signs in the diagonal of $\bs D$ out of $P=N+\nu$. A Gaussian with standard deviation $\sqrt{3N/2}$ centered at $-N/2$ is plotted for the sake of comparison with the $k=1$ data. }
  \label{fig:min-eigs}
\end{figure}

This generalizes to $k<P$. The left edge of the full eigenvalue distribution for an ensemble of $\hat{\bs H}$ generated from random $\hat{\mathbfcal{Q}}$ using $\bs D$ with fixed $k<P$ negative signs converges to $\langle\lambda_-\rangle=-c(k) N$, where $c(k)$ is an order one number less than 4. The scaling of the width of the distribution of $\lambda_{-}$  with $N$ is also suppressed with respect to the mean $\langle \lambda_{-}\rangle$, just as for the Tracy-Widom distribution describing the $k=0$ and $k=P$ cases. For example, when $k=1$, $\langle |\lambda_{-}|\rangle=N/2$ and the standard deviation  is $\sqrt{3 N/2}$. A sample of the $\lambda_{-}$ distributions for varying $k$ is shown in Figure \ref{fig:min-eigs}.  

Therefore we see that for all $k$ and at large $N$, the standard deviation in the smallest (most negative) eigenvalue is much smaller than its mean.

\subsection{Face neighbor distances} \label{appendix-distances}
Each face neighbor  is specified by the signed unit-normal vector to the face it shares with vacuum $A$'s cube,
\begin{equation}
\pm\bs e^{(J)} =  (0,0,...,0,\pm1,0,...,0) \label{face vectors} \,.
\end{equation}
In canonical coordinates the face neighbor vacuum's location is well-approximated by
\begin{equation}
 \bs \Theta_\text{B} \approx  \bs \Theta_\text{A}\pm 2\pi  (\bold Q^\top \bold Q)^{-1} \bold Q^\top \bs e^{(J)} \,. \label{parent-location}
\end{equation}
With the hatted  notation introduced in \S \ref{appendix-hessian}, the mean of the distribution of the face neighbor distances $\lVert \bs \Theta_{\text A} - \bs \Theta_{\text B} \rVert_2$ is
\begin{equation} \label{PDdistance}
\langle \lVert \bs \Theta_{\text A} - \bs \Theta_{\text B} \rVert_2 \rangle = \frac{2 \pi f}{\sigma_\mathcal{Q}} \times \langle \lVert (\hat{\mathbfcal Q}^\top \hat{\mathbfcal Q})^{-1} \hat{\mathbfcal Q}^\top \bs e^{(J)} \lVert_2 \rangle \,.
\end{equation}
A non-obvious fact is that the hatted quantity on the lefthand side of (\ref{PDdistance}) turns out to be order one at large $N$,
\begin{equation} \label{numerics1}
 \langle \lVert (\hat{\mathbfcal Q}^\top \hat{\mathbfcal Q})^{-1} \hat{\mathbfcal Q}^\top \bs e^{(J)} \rVert_2 \rangle = \mathcal{O}(1) \,.
\end{equation}
The resulting canonical field distances between face neighbors is
\begin{equation} \label{PDdistance-num}
\langle \lVert \bs \Theta_{\text A} - \bs \Theta_{\text B} \rVert_2 \rangle \approx \frac{2 \pi f}{\sigma_\mathcal{Q}} \times \mathcal{O}(1) \,.
\end{equation}
Note the face neighbor distance is suppressed by a factor $N$ relative to the tile diameters, which go like $N$ when $\sigma_\mathcal{Q} \sim {\cal{O}}(1)$,  and $N^{3/2}$ in the sparse charge matrix case where $\sigma_\mathcal{Q} =\mathcal{O}(1)/\sqrt{N}$.

\begin{figure}
  \centering
  \includegraphics[width=0.83\linewidth]{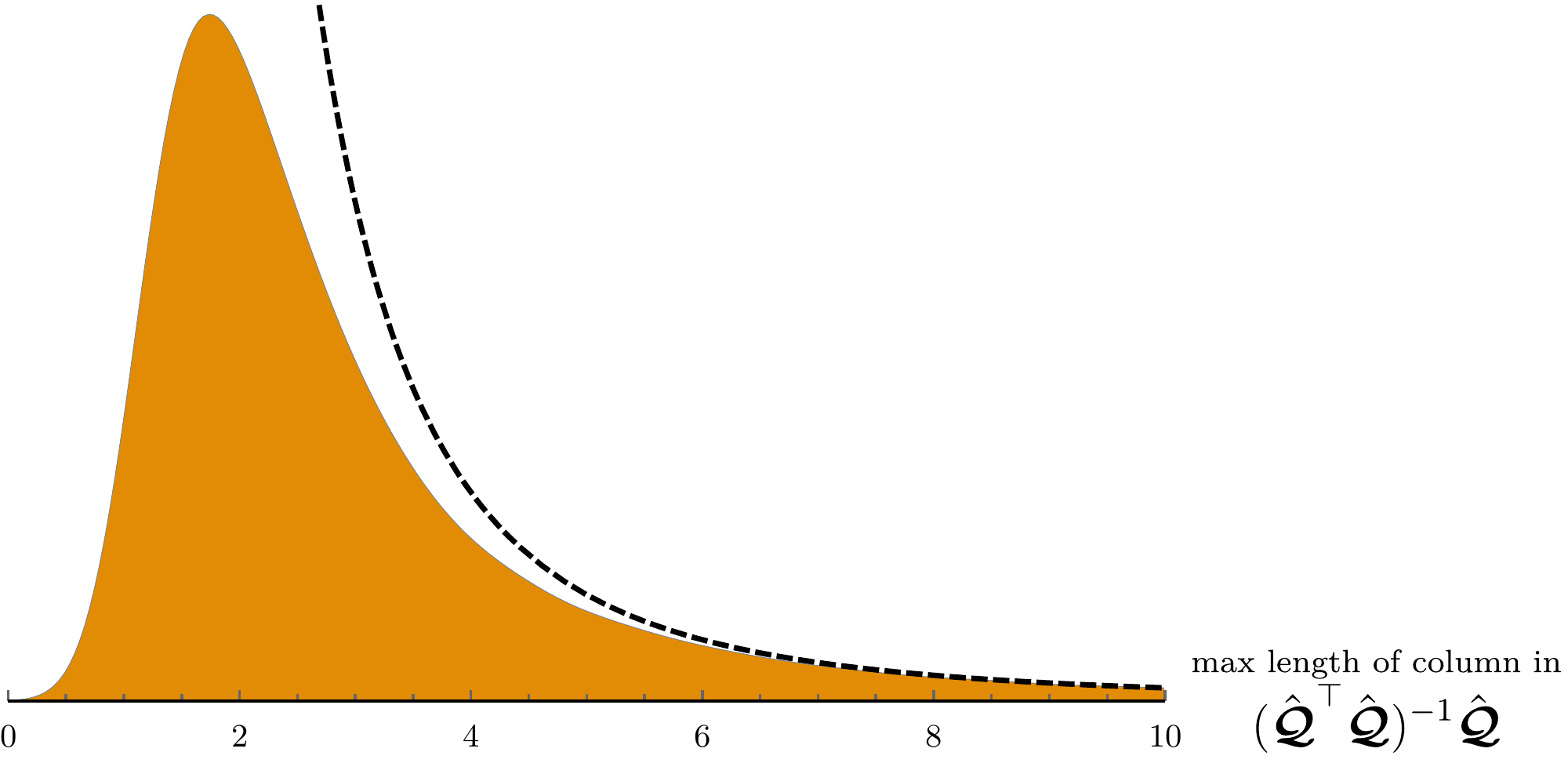}
  \caption{\small Numerical probability density of the maximum $2$-norm length of the columns of the matrix $(\hat{\mathbfcal Q}^\top \hat{\mathbfcal Q})^{-1} \hat{\mathbfcal Q}^\top$, where the entries of $\hat{\mathbfcal Q}$ are i.i.d. according to $\mathcal{N}(0,1)$. This  quantity appears in an estimate for the field space distance between face neighbor minima.  In this example $N=200, P=201$, and we have fit the tail of the distribution to the form $\text{(length)}^{-f(1)}$, with $f(1) \approx 3$.  \label{parent-target-histogram}}
\end{figure}

An important aspect of the distribution of $\lVert \bs \Theta_{\text A} - \bs \Theta_{\text B} \rVert_2$, and of the distribution of maximum distances, is that it appears to be heavy-tailed for $\nu\ll N$. Taking the asymptotic form  $ \lVert \bs \Theta_{\text A} - \bs \Theta_{\text B} \rVert_2^{-f(\nu)}$ for the tail, numerics suggest

\begin{equation} \label{f(nu)guessxxx}
	f(\nu) = 2 + \nu \,,
\end{equation}
although we have no analytic argument for this relation. A numerical PDF of the maximum distance is shown in Figure \ref{parent-target-histogram}.
Though lacking a derivation of the fall-off exponent $f(\nu)$, the basic fact that the distribution is heavy-tailed at all can be understood heuristically. Moreover, this reasoning simultaneously explains the $\mathcal{O}(1)$ expectation for the hatted distance between face neighbors (the quantity on the left-hand side of (\ref{numerics1})).

 Consider first the case when $P=N$, so the charge matrix is square and invertible. Then $( \hat{\mathbfcal{Q}}^\top \hat{\mathbfcal{Q}})^{-1}\hat{\mathbfcal{Q}}^\top$ reduces to $\hat{\mathbfcal{Q}}^{-1}$, so we recognize the hatted distance between two face neighboring minima
\begin{equation}
 \lVert (\hat{\mathbfcal Q}^\top \hat{\mathbfcal Q})^{-1} \hat{\mathbfcal Q}^\top \bs e^{(J)} \rVert_2 \label{hat-distance}
\end{equation}
 as the length of the $J$th column of $\hat{\mathbfcal{Q}}^{-1}$. 

When $P>N$ the matrix $\hat{\mathbfcal{Q}}$ of course is no longer invertible, but the rectangular matrix $( \hat{\mathbfcal{Q}}^\top \hat{\mathbfcal{Q}})^{-1}\hat{\mathbfcal{Q}}^\top$ is -- in a precise sense -- the closest thing to the inverse of $\hat{\mathbfcal{Q}}$. Start with a singular value decomposition of $\hat{\mathbfcal{Q}}$, 
\begin{equation}
\hat{\mathbfcal{Q}}=\bold V \bold \Sigma \bold U^\top \,.
\end{equation}
Here $\bold U$ and $\bold V$ are orthogonal $N\times N$ and $P\times P$ matrices respectively, and the $P\times N$ matrix $\bs \Sigma$ contains the singular values of $\hat{\mathbfcal{Q}}$ along the diagonal of its top $N \times N$ diagonal subblock, followed by $P-N$ zero rows. In other words $\bold \Sigma$ has the form,
\begin{equation}
\bold{\Sigma}=\begin{bmatrix}\sigma_1&0&\dots&0\\
0&\sigma_2&\dots&0\\
\vdots&&\ddots&\vdots\\
0&0&\dots&\sigma_N\\
0&0&\dots &0\\
\vdots&\vdots&&\vdots\\
0&0&\dots &0
\end{bmatrix} \,.
\end{equation}
 The first $N$ columns of $\bold V$ form an orthonormal basis for the constraint surface $\Sigma$ when it is embedded in $\mathbb{R}^P$, while $\bold V$'s remaining $P-N$ columns do so for the orthogonal complement.

To simplify notation we label the inverse matrix as $\bold A\equiv(\hat{\mathbfcal{Q}}^\top \hat{\mathbfcal{Q}})^{-1}$. Note that the eigenvectors of $\bold A$ are the right-singular vectors of $\hat{\mathbfcal{Q}}$ (the columns of $\bold U$, which we'll label as $\bold u_i$, and similarly for $\bold V$). The eigenvalues of $\bold A$ are $a_i=1/\sigma_i^2$. While $\hat{\mathbfcal{Q}}$ maps the unit vector ${\bold u}_i$ to the (generally not normalized) $P$-vector $\sigma_i \bold{v}_i$, the matrix $\bold A \hat{\mathbfcal{Q}}^\top$ maps ${\bold v}_i$ to $1/\sigma_i \bold{u}_i$. This is the sense in which the pair $\bold A \hat{\mathbfcal{Q}}^\top$ and $\hat{\mathbfcal{Q}}$ can be thought of as the rectangular analog of a matrix and its inverse. The rectangularity is reflected in the fact that ${\bold v}_i$ plus any $P$-vector in $\Sigma^\perp$ still maps to $1/\sigma_i \bold{u}_i$. 

Going forward we also suppress the $(J)$ superscript on the face vector. To develop an expectation for the magnitude of the lengths $\lVert \bold A \hat{\mathbfcal{Q}}^\top \bs e \rVert_2$, start by expanding $\hat{\mathbfcal{Q}}^\top\bs e$ in the right-singular vectors of $\hat{\mathbfcal{Q}}$,
\begin{equation}
\hat{\mathbfcal{Q}}^\top \bs e=\hat{\mathbfcal{Q}}^\top \bs e_\parallel=\sum_{i=1}^N b_i \bold{u}_i \,.
\end{equation}
Then
\begin{equation}
\bold A \hat{\mathbfcal{Q}}^\top \bs e = \sum_{i=1}^N \frac{b_i}{\sigma_i^2} \bold{u}_i \,.
\end{equation}
Since $\bold U$ is orthogonal, the length of $\bold A \hat{\mathbfcal{Q}}^\top \bs e$ is the square root of the sum of the squares of the expansion coefficients,
\begin{equation}
\lVert \bold A \hat{\mathbfcal{Q}}^\top \bs e \rVert_2 = \sqrt{\sum_{i=1}^N \frac{b_i^{~2}}{\sigma_i^{~4}}} \,.
\end{equation}
The spectrum of $\bold A$ is known because it is an inverse Wishart matrix ($\hat{\mathbfcal{Q}}$ is a Wigner matrix).  The task then amounts to correctly estimating the expansion coefficients $b_i$. It might be tempting to think these would be roughly equal, or, perhaps a bit more carefully, that they be normally distributed with approximately equal variances. This is not the case, however.

The $b_i$ are strongly correlated with the ${a_i}$ due to the fact that $\bs e$ is normalized and $\hat{\mathbfcal{Q}} \bold A \hat{\mathbfcal{Q}}^\top$ is an orthogonal projector. Qualitatively, the components of $\bold A\hat{\mathbfcal{Q}}^\top \bs e$ in large-eigenvalued eigendirections must be small enough that their image under $\hat{\mathbfcal{Q}}$ have norm $\leq 1$.
The correlation can be understood precisely by relating the $b_i$ to the expansion coefficients of $\bs e_\parallel$ in the \emph{left}-singular vectors ($\bs v_i$), which we'll denote by $\beta_i$,
\begin{equation}
\bs e_\parallel=\sum_{i=1}^N \beta_i \bs v_i \,.
\end{equation}
In light of the fact that $\bold A \hat{\mathbfcal{Q}}^\top:{\bold v}_i \mapsto 1/\sigma_i \bold{u}_i$ the expansion coefficients are related by $\beta_i=b_i/\sigma_i $. The $\beta_i^2$ must sum to a value $\leq 1$ because $\bs e$ is normalized. Recall that face vectors generically have $\lVert \bs e_\parallel \rVert_2 \approx1$ and $\bs e_\perp\ll 1$ for $P-N\ll N$ because $\Sigma$ is randomly oriented. Likewise, the projection onto $\Sigma$ itself, $\bs e_{\parallel}$,  bears no special relation to the orthogonal directions $\bold v_i$. Consequently, we expect values for \emph{these} expansion coefficients -- the $\beta_i$ -- to be comparable to one-another. They are the components of a normalized vector so the magnitude to expect is $\beta_i \sim 1/\sqrt{N}$ since,
\begin{equation}
\sum_{i=1}^N\beta_i^2\approx 1 ~~ \Rightarrow ~~ N \beta_i^2 \sim 1 \,. \label{typical-alpha}
\end{equation}
A rough estimate of the mean/median of $\lVert \bold A \hat{\mathbfcal{Q}}^\top \bs e \rVert_2$ can be obtained setting all $\beta_i^2=1/N$. One then concludes a value of $\langle \lVert \bold A \hat{\mathbfcal{Q}}^\top \bs e \rVert_2 \rangle\approx \langle \sqrt{\lambda_\text{A}}\rangle$, where $\lambda_\text{A}$ represents a blindly-drawn eigenvalue of the matrices in an inverse Wishart ensemble\footnote{ It should be noted that while $1/\sqrt{N}$ is the magnitude to expect for the $\beta_i$, the upper bound on any given $|\beta_i|$ is of course still $1$. In terms of the $|b_i|$ this translates to a strict boundary at $|b_i|=1/\sqrt{{a_i}}$. We have confirmed numerically that a marked decrease in the $|b_i|$ is observed at $3/\sqrt{N a_i}$, which is consistent with having gaussian distributed $\beta_i$ with variance $1/N$.}. Numerically we find the median of this is $\mathcal{O}(1)$. This is (literally) the observation (\ref{numerics1}).

Furthermore, the eigenvalue distribution of the inverse Wishart ensemble is heavy-tailed.  This explains the  heavy tail of the PDF in Figure \ref{parent-target-histogram}, albeit without an analytic estimate of the numerically observed fall-off rate $f(\nu)$. An important conclusion  is that parent-to-target distances much larger than the median (\ref{PDdistance}) are much more frequent than distances much smaller than (\ref{PDdistance}) (which appear to be exponentially suppressed). 

\section{Refined stability bound} \label{appendix-stability}
The purpose of this appendix is provide a more detailed version of the stability analysis performed in \S\ref{sec:vactransitions} for tunneling from minima with nearly zero vacuum energy. We also account for the variations among the neighbors of a given degree-$k$ set, which is currently ignored in the estimate of an upper bound on thin-wall tunneling exponent $B$ in \S\ref{sec:vactransitions}. The result will be a more precise bound. We also study the distribution of vacuum energy differences and show that this may improve the bound on $B$ for tunneling to face neighbors.

\subsection{Stability}
The set of degree-$k$ neighbors of a given parent are not identical. They vary both in the widths of the barriers separating them from the parent, and in their difference in vacuum energy density compared to the parent. The worst case scenario from the perspective of a parent's stability, is when the degree-$k$ neighbor with thin-wall tension ($\sigma$) fluctuated the most toward small values is \emph{also} that with largest vacuum energy difference, $\epsilon$.  Any other correlation between the two quantities would serve to increase $B$, and render the vacuum more stable.

Recall that 
\begin{equation}\label{sigma-bump-eq}
\sigma \approx {2\pi k \Lambda^4 \over \sqrt{|V_*''|}} \,,
\end{equation}
with $|V_*''|$ well-approximated by
\begin{equation}
|V_*''| =|\lambda_{-}| \frac{\sigma_\mathcal{Q}^2 \Lambda^4}{f^2} \,,
\end{equation}
and where $\lambda_{-}$ is the most negative eigenvalue of $ \hat{\bs H}$. As shown in appendix section \ref{appendix-hessian}, the mean of the distribution of the most negative eigenvalues in an $\hat{\bs H}$-ensemble behaves at large $N$ as 
\begin{equation}
	\langle\lambda_{-}\rangle=-c(k) N \,,
\end{equation}
where $c(k)$ is an $\mathcal O(1)$ constant, ranging from $c(1) \approx 1/2$, to $c(P) \approx 4$.
Although using $\langle \lambda_{-}\rangle$ in (\ref{sigma-bump-eq}) captures the contribution to the tension from the near-saddle region for a \emph{typical} degree-$k$ neighbor, what matters is the extremal neighbor; the leftmost outlier $\lambda_-$ among the set of about $2^k {P\choose{k}}$ neighbors. To account for this we'll write $\lambda_{\text{edge}}=C_1\langle\lambda_{-}\rangle$. The standard deviation of the distribution of $\lambda_{-}$ goes like $\sqrt{3N/2}$, which is suppressed by a power of $N^{1/2}$ relative to the mean, implying $C_1$ will tend to $1$ as $N\rightarrow \infty$. However, for $N=500$, a value of about $C_1=1.5$ is appropriate. Putting things together, we have
\begin{equation}
B_{\text{degree-}k} \simgeq \frac{6^3 \pi^6}{C_1^2} \frac{ k^4}{c^2(k)}\frac{\Lambda^8 f^4}{\sigma_{\mathcal{Q}}^4N^2\epsilon^3} \,
\end{equation}

An \emph{upper} bound on $\epsilon=|\Delta V_{\text{lowest outlier}}|$ completes the calculation. For tunneling from a minimum with zero vacuum energy one has $\epsilon<|V_0|$, since the global minimum has energy $V_\text{global min}\geq -|V_0|$.  Plugging in $\epsilon = |V_0|$   reproduces the bound  in \S\ref{sec:vactransitions}, (\ref{boundsummary}) provided one sets $C_1=1$ and $c(k)=c(1)=1/2$ (since the estimate there ignores fluctuations in $\lambda_-$ and the sublinear $k$-dependence of the random matrix theory coefficient $c(k)$).

It turns out that face neighbors have smaller $\Delta V$ outliers than $-|V_0|$ in certain parameter regimes. This is discussed in detail in the following section. For example, for $V_0=\Lambda^4$ and $P=500$ the net effect is about a factor of a half, $\epsilon_{k=1}<|V_0|/2$.  If the decay rate  is dominated by tunneling to these neighbors (as we expect), this factor of 2  contributes roughly an order of magnitude increase to $B$, since it enters via $1/\epsilon^3$.

\subsection{Vacuum energy differences}\label{appendix-deltaVs}
The vacuum energy difference between two neighbors can be approximated by a simple Taylor expansion of
\begin{equation}
V_{\text{B}}=V_{\text{aux}}(\bold Q\bs\Theta_{\text{B}})\approx V_{\text{aux}}(\bold Q\bs\Theta_{\text{A}}+2\pi\Po\bs v_k) \,.
\end{equation}
The result to second order in $\lVert \bs v_k^\perp \rVert_2  \equiv \lVert  \Pob \bs v_k \rVert_2$ is
\begin{equation}
\Delta V=V_{\text{B}}-V_{\text{A}}=\Lambda^4\left(2\pi \sqrt{2\mu P} \lVert \bs v_k^\perp \rVert_2 \cos(\psi)+2\pi^2 \lVert \bs v_k^\perp \rVert_2^2\right) \,, \label{deltaV-expansion}
\end{equation}
where $\psi$ is the angle between $\bs n_{\text{A}}^\perp$ and $\bs v_k^\perp$, and $\mu = 2\pi^2 \lVert \bs n_\text{A}^\perp \rVert_2^2 /P$. The derivation of (\ref{deltaV-expansion}) is included at the end of this section for completeness.

The quadratic contribution to (\ref{deltaV-expansion}) is positive definite. Since the $\bs v_k$ come in pairs that are equal in magnitude but exactly opposite in direction\footnote{Because for every nontrivial entry there is a choice of $\pm 1$.},  half of the degree-$k$ neighbors have $\cos(\psi)>0$, and half have $\cos(\psi)<0$. In any given model, parent vacua with sufficiently high $\mu$ have first order contributions to the $\Delta V$'s that tend to dominate the quadratic ones for most of the neighbors in a given degree-$k$ set, resulting in the parent's set of $\Delta V$'s being approximately symmetric about zero, for each $k$. As a parent's $\mu$ value decreases though, this balance is thrown off because an increasing number of negative first order contributions -- which are proportional to $\sqrt{\mu}$ -- will be partially canceled, if not entirely overwhelmed, by the positive quadratic term. 

Whether positive or negative, the $\Delta V$ to a parent's lowest energy neighbor can be estimated with relative ease because of the observation that the two-norms of the $\bs v^\perp_k$ and their orientations with respect to the $\bs n_\text{A}^\perp$  appear to be uncorrelated in random axion landscapes. These are the only two sources of variability in a parent's $\Delta V$'s for fixed $k$.\footnote{This applies to all neighbors for which the second order Taylor expansion (\ref{deltaV-expansion}) is accurate. Since we are only interested in tunneling to lower energy neighbors, approximating the vacuum energy of the neighbor by $\Lambda^4 2\pi^2 d_{\text{B}}^2$ is even more accurate than the analogous expression is for the parent. The difference between these two quadratic approximations to the vacuum energy directly gives (\ref{deltaV-expansion}) without further assumption.} When $\nu \gtrapprox 10$ the $\cos(\psi)$ behave like the dot products between a set delocalized unit vectors in $\nu$ dimensions, implying that the $\cos(\psi)$ are approximately normal distributed with mean zero and standard deviation $1/\sqrt{\nu}$. The fall-off eventually deviates from the Gaussian, becoming sharper due to the fact that $|\cos(\psi)|\leq 1$. For a sample of different low CC parents, the distribution of maximum dot product factors -- one $\text{Max}(\cos(\psi))$ from each parent taken over its fixed $k$ neighbors-- is peaked at $x^*/\sqrt{\nu}$ with $x^*$ ranging from about $2.2$ to $2.9$ for $\nu=10$ to $\nu=15$ and $P=400$.

Now for the two-norms; the mean of the distribution of the set of $\lVert \bs v_k^\perp \rVert_2$ for fixed $k$ is in accordance with the naive prediction based on assuming $P$ independent normal distributed entries with standard deviation $\sqrt{\nu k}/P$, namely the mean of a $\chi$-distribution in $P$ variables scaled by the standard deviation of the individual entries, i.e.
\begin{equation}
\langle \lVert \bs v_k^\perp \rVert_2 \rangle = \sqrt{\nu k/P}  \label{mean-vkperp}
\end{equation} 
(cf.~(\ref{vp})). Though the means agree, the actual $\{ \lVert \bs v_k^\perp \rVert_2 \}$ are more widely distributed.\footnote{The spread of an actual set of $\lVert \bs v_k^\perp \rVert_2$ is wider than the naive expectation  by a factor of $5$ or so,
\begin{align*}
\text{Std. Dev. (scaled }\chi)&=\sqrt{ \frac{\nu k}{P} }\sqrt{P-2\left(\frac{\Gamma((P+1)/2)}{\Gamma(P/2)}\right)^2}\\
&\sim \sqrt{ \frac{2\nu k}{P}} \,, \\
\text{Std. Dev. (actual } \lVert \bs v_k^\perp \rVert_2 )&\approx 5 \sqrt{ \frac{2\nu k}{P} } \,.
\end{align*}
The reason for this discrepancy is that the entires of $\Pob$ are correlated.}
 
With these considerations, the $\Delta V$ for a general degree-$k$ neighbor can be expressed in terms of the natural scales by setting $|\cos(\psi)|=x/\sqrt{\nu}$ and $\lVert \bs v_k^\perp \rVert_2 = y \sqrt{\nu k/P}$ in equation (\ref{deltaV-expansion}). We also take the parent's $\mu$ to  be $\mu_{\text{low CC}}\approx V_0/(P\Lambda^4)$ from now on. The result, expressed as a function of the dimensionless order $1$ quantities $x$ and $y$ that entirely capture the variation in $\Delta V$ across neighbors of fixed $k$, is
 \begin{equation}
\frac{\Delta V}{\Lambda^4}=\pm2\pi\sqrt{2 V_0/\Lambda^4}~x y \sqrt{\frac{k}{P}}+2\pi^2~ y^2\frac{\nu k}{P} \,, \label{general-deltaV-neighbor}
\end{equation}
where the $+$ applies to the half with $\cos(\psi)>0$ and $-$ to those with $\cos(\psi)<0$, and we bear in mind that $0\leq  x\leq\sqrt{\nu}$ and $0\leq y \leq \sqrt{P/( \nu k)}$. 

The optimal conditions for making $\Delta V$ as negative as possible come from $x=\sqrt{\nu}$ (and obviously selecting the $-$ sign), and $y$ given by 
\begin{equation}
y_{\text{optimal}}= \frac{\sqrt{ P V_0/\Lambda^4}}{\pi \sqrt{2 k\nu}} \,.
\end{equation}
Evaluating $\Delta V(x,y)$ at the optimal values turns out to simply give $-V_0$, the energy gap to the global minimum. At higher $k$ one can expect to saturate the inequality $\Delta V > -V_0$ because there are exponentially many neighbors, but for low $k$ the ${2^k}{ {P}\choose{k}}$ draws of $(x,y)$ may be insufficient for ensuring that at least one pair of the $(x,y)$ is close to $(x_{\text{optimal}},y_{\text{optimal}})$.

Without an analytic expression for the PDF governing the $\{ x_i\}$ and $\{ y_i\}$ though, the only way to determine the outlier $(x_i,y_i)$ at low $k$ is numerically. For $\nu= 10$ to about $15$ and $P=250$--$500$ we find that the the aforementioned value of $x^*$ together with $y^*=1$  yields a good approximation to the $\Delta V_\text{outlier}$ for the special case of $k=1$. Essentially, this is because the wider of the two distributions controls the outlier value if there are not enough draws to densely sample the allowed $\Delta V$'s, coupled to the fact that the optimal $y$ happens to be relatively close to $1$ (unlike for high $k$, where it approaches zero). Hence for $k=1$ we may apply the estimate
\begin{equation}
\frac{\Delta V_{\text{outlier, }k=1}}{\Lambda^4}=-2\pi\sqrt{2 V_0/\Lambda^4}~\frac{x^*(\nu,P)}{\sqrt{P}}+2\pi^2~ \frac{\nu }{P} \label{kis1-outlier}
\end{equation}
 and define 
\begin{equation}
C_3(V_0,\nu,P)=  2\pi \left(-\sqrt{2 V_0/\Lambda^4}~x^* +\frac{\pi \nu }{\sqrt{P}}\right) \,, \label{C3-kis1}
\end{equation}
so that
\begin{equation}\label{C3-bound}
\Delta V_{\text{outlier, }k=1}=\frac{C_3 \Lambda^4}{\sqrt{ P }} \,.
\end{equation}
For $V_0=\Lambda^4$ and $P=500$  this results in a value of about $ -12$ for $C_3$.

As $k$ increases there is a transition from an outlier energy difference of $\epsilon=|C_3|  \Lambda^4\sqrt{k/P} $ for $k=1$ to an outlier with $\epsilon=V_0$. In this parameter regime it turns out that the transition occurs immediately at $k=2$. The particular values here result in an outlier difference for $k=1$ that is about half as large as the global bound. Accounting for such order one factors at $V_0=\mathcal{O}(1)\Lambda^4$ can be significant because they enter with a third power in the denominator of $B$, and so are capable of increasing the numerical factor in the bound by orders of magnitude.  

Expressing the outlier $\Delta V$ as $\sim 1/\sqrt{P}$, as we've done in (\ref{C3-bound}), is useful in that it accurately reflects the $N$ dependence for larger $\mu$ parents (where the linear term in $\Delta V$ dominates). However, writing the outlier $\Delta V$ in this manner obscures the fact that the absolute minimum value it ever takes on is $-|V_0|$. This fact is encoded in the definition of $C_3$ in (\ref{C3-kis1}), but the property that the refined bound on $\Delta V$ is always stronger than the global bound can be made manifest as follows. Evaluating (\ref{general-deltaV-neighbor}) at the outlier values for $x$ and $y$ expressed in terms of the optimal ones:  $x^*=r_x x_{\text{optimal}}$ and $y^*=r_y y_{\text{optimal}}$, results in the expression,
\begin{equation}
\Delta V_{\text{outlier}}= (r_y^2-2 r_x r_y)V_0 \,.
\end{equation}

By writing the outlier energy gap between quadratic domain vacua as a fraction times $V_0$, one can immediately see the degree to which using the refined bound on $\Delta V$ over the global one improves the lower bound on the tunneling exponent $B$. The benefit amounts to an enhancement of $B_{\text{min}}$ by the inverse of the fraction, cubed. An expansive numerical study the typical values of $r_x$ and $r_y$ in different parameter regimes would enable one to determine the typical benefit of employing the refined bound for very low $k$ neighbors (possibly only $k=1$, as the numerical survey may reveal that the transition to saturation never occurs above $k=2$). 

Finally, we include the derivation of the expression we began with, (\ref{deltaV-expansion}). Start by evaluating the potential at the leading order locations for the two vacua:
\begin{align*}
\Delta V&=V(\bs\Theta_{\text{B}})-V(\bs\Theta_{\text{A}})\\
&\approx V_{\text{aux}}(\bold Q\bs\Theta_{\text{A}}+2\pi\Po \bs v_{k})-V_{\text{aux}}(\bold Q\bs\Theta_{\text{A}})\\
&=\Lambda^4\sum_{J=1}^P 1 - \cos \left[ 2\pi(\bs n_{\text{A}}+\bs v_{k}-\bs n^\perp_{\text{A}} - \bs v_k^\perp)^J \right] - \left( 1-\cos \left[ 2\pi ( \bs n_{\text{A}}-\bs n^\perp_{\text{A}})^J \right] \right) \\
&=\Lambda^4\sum_{J=1}^P 1-\cos \left[ 2\pi(\bs n_{\text{A}}-\bs n^\perp_{\text{A}}-\bs v_k^\perp)^J \right] -\left( 1-\cos \left[ 2\pi(\bs n_{\text{A}}-\bs n^\perp_{\text{A}})^J \right] \right) \,.
\end{align*}
Now define $\epsilon^J\equiv 2\pi (\bs v_k^\perp)^J$ and expand the first cosine term in the sum about the argument of the second, i.e. expand in $\epsilon^J$. The result is
\begin{equation}
\Delta V=\Lambda^4\sum_{J=1}^P\left(-\sin[2\pi(\bs n_{\text{A}}-\bs n^\perp_{\text{A}})^J]\epsilon^J+\frac{|\epsilon^J|^2}{2}\cos[2\pi(\bs n_{\text{A}}-\bs n^\perp_{\text{A}})^J]\right)+\mathcal{O}[(\epsilon^J)^3] \,.
\end{equation}
Now, further expand the sine and cosine about $(2\pi \bs n_\text{A})^J$,
\begin{equation}
\Delta V=\Lambda^4\sum_{J=1}^P\left(2\pi (n^{\perp}_{\text{A}})^J \epsilon^J+(1-(n^{\perp J}_{\text{A}})^2 )~\frac{|\epsilon^J|^2}{2}\right)+\mathcal{O}[(\epsilon^J)^3] \,.
\end{equation}
Using $(n_{\text{A}}^{\perp })^J \ll 1$, we arrive at the following simple expression for the energy difference between quadratic domain neighbors:
\begin{equation}
\Delta V = 2 \pi^2 \Lambda^4 \left(2 \bs n^\perp_{\text{A}}\cdot \bs v_k^\perp + \lVert \bs v_k^\perp \rVert_2^2 \right) \,,
\end{equation}
which we can recognize as  the difference in the parent and neighbor lattice point's two-norm distances to the constraint surface, times $2\pi^2$ (as it had to be). Plugging in $\lVert \bs n^\perp_{\text{A}} \rVert_2  = \sqrt{\mu_{\text{A}}P / 2\pi^2}$ and defining $\psi$ as the angle between $\bs n^\perp_{\text{A}}$ and $\bs v_k^\perp$ gives equation (\ref{deltaV-expansion}).

\bibliographystyle{klebphys2}
\bibliography{tunnelingrefs}
\end{document}